\documentclass[nohyper]{JHEP3}

\usepackage{amsmath,amssymb}
\usepackage{graphicx}
\usepackage{cite}

\newcommand{\la}{\left\langle}
\newcommand{\ra}{\right\rangle}
\newcommand{\lc}{\left[}
\newcommand{\rc}{\right]}
\newcommand{\lp}{\left(}
\newcommand{\rp}{\right)}

\newcommand{\mrexp}{\mathrm{exp}}
\newcommand{\dat}{\mathrm{dat}}
\newcommand{\art}{\mathrm{art}}
\newcommand{\rep}{\mathrm{rep}}
\newcommand{\net}{\mathrm{net}}

\newcommand{\cor}{\mathrm{cor}}
\newcommand{\parr}{\mathrm{par}}
\newcommand{\stat}{\mathrm{stat}}

\newcommand{\tot}{\mathrm{tot}}
\newcommand{\cov}{\mathrm{cov}}
\newcommand{\NNe}{\mathrm{NN}\lp E_\nu\rp}
\newcommand{\cz}{c_\nu}
\newcommand{\draft}[1]{}

\newcommand{\Efin}{E^\text{fin}}
\newcommand{\Emin}{E_\text{min}}

\title{Determination of the Atmospheric Neutrino Fluxes from
  Atmospheric Neutrino Data}

\author{M.~C.~Gonzalez-Garcia\\
  C.N. Yang Institute for Theoretical Physics,\\
  State University of New York at Stony Brook,\\
  Stony Brook, NY 11794-3840, USA\\
  {\rm and:} Instituto de F\'\i sica Corpuscular,
  Universitat de Val\`encia -- C.S.I.C.,\\
  Edificio Institutos de Paterna, Apt 22085, E-46071 Val\`encia, Spain\\
  {\rm after October 1st:}
  Instituci\'o Catalana de Recerca i Estudis Avan\c{c}ats (ICREA), \\
  Departament d'Estructura i Constituents de la Mat\`eria,
  Universitat de Barcelona,\\
  Diagonal 647, E-08028 Barcelona, Spain\\
  E-mail: \email{concha@insti.physics.sunysb.edu}}
\author{M.~Maltoni\\
  The Abdus Salam International Centre for Theoretical Physics (ICTP),\\
  Strada Costiera 11, I-31014 Trieste, Italy\\
  {\rm and:} Departamento de F\'isica Te\'orica C-XI,
  Universidad Aut\'onoma de Madrid,\\
  Cantoblanco, E-28049 Madrid, Spain\\
  E-mail: \email{maltoni@delta.ft.uam.es}}
\author{J.~Rojo\\
  Departament d'Estructura i Constituents de la Mat\`eria,
  Universitat de Barcelona,\\
  Diagonal 647, E-08028 Barcelona, Spain\\
  {\rm after October 1st:}
  LPTHE, CNRS UMR 7589, Universit\'e P.\ et M.\ Curie (Paris VI),\\
  Universit\'e Denis Diderot (Paris VII), 75252 Paris Cedex 05, France\\
  E-mail: \email{joanrojo@ecm.ub.es}}


\abstract{%
  The precise knowledge of the atmospheric neutrino fluxes is a key
  ingredient in the interpretation of the results from any atmospheric
  neutrino experiment.  In the standard data analysis, these fluxes
  are theoretical inputs obtained from sophisticated numerical
  calculations based on the convolution of the primary cosmic ray
  spectrum with the expected yield of neutrinos per incident cosmic
  ray. In this work we present an alternative approach to the
  determination of the atmospheric neutrino fluxes based on the direct
  extraction from the experimental data on neutrino event rates. The
  extraction is achieved by means of a combination of artificial
  neural networks as interpolants and Monte Carlo methods for faithful
  error estimation.}

\preprint{%
  hep-ph/0607324\\
  UB-ECM-PF-06/16\\
  YITP-SB-06-31\\
  IC/2006/064}

\begin{document}

\section{Introduction and Motivation}
One of the most important breakthroughs in particle physics, and the
only solid evidence for physics beyond the Standard Model, is the
discovery -- following a variety of independent
experiments~\cite{neutrev2} -- that neutrinos are massive and
consequently can oscillate among their different flavor eigenstates.
The flavour oscillation hypothesis has been supported by an impressive
wealth of experimental data, one of the most important pieces of
evidence coming from atmospheric
neutrinos~\cite{atmexpold,sk,macro,soudan2}.

Atmospheric neutrinos originate in the collisions of cosmic rays with
air nuclei in the Earth's atmosphere. The collision produces mostly
pions (and some kaons), which subsequently decay into electron and
muon neutrinos and anti-neutrinos.  These neutrinos are observed in
underground experiments using different
techniques~\cite{atmexpold,sk,macro,soudan2}. In particular, in the
last ten years high precision and large statistics data has been
available from the Super-Kamiokande experiment~\cite{sk}, which has
clearly established the existence of a deficit in the $\mu$-like
atmospheric events with the expected distance and energy dependence
from $\nu_\mu\rightarrow \nu_\tau$ oscillations with oscillation
parameters $\Delta m^2_{\mathrm{atm}} \sim 2\times 10^{-3}$ eV$^2$ and
$\tan^2\theta_{\mathrm{atm}}=1$.  This evidence has also been
confirmed by other atmospheric experiments such as MACRO~\cite{macro}
and Soudan 2~\cite{soudan2}.

The expected number of atmospheric neutrino events depends on a
variety of components: the atmospheric neutrino fluxes, the neutrino
oscillation parameters and the neutrino-nucleus interaction cross
section. Since the main focus of atmospheric neutrino data
interpretation has been the determination of neutrino oscillation
parameters, in the standard analysis the remaining components of the
event rate computation are inputs taken from other sources. In
particular, the fluxes of atmospheric neutrinos are taken from the
results of numerical calculations, such as those of
Refs.~\cite{honda,bartol,others}, which are based on the convolution
of the primary cosmic ray spectrum with the expected yield of
neutrinos per incident cosmic ray~\cite{fluxrev}.

The oscillations of $\nu_\mu$ can also be tested in Long Baseline
(LBL) experiments, using as neutrino source a controlled beam of
accelerator neutrinos.  The results of the first two of these LBL
experiments, K2K~\cite{K2K} and MINOS~\cite{MINOS}, confirm, both in
the observed deficit of events and in their energy dependence, that
accelerator $\nu_\mu$ oscillate as expected from oscillations with the
parameters inferred from the atmospheric neutrino data.  Furthermore
they already provide a competitive independent determination of the
relevant $\Delta m^2$, and with time either MINOS and/or other future
LBL experiments~\cite{T2K,FutureLBL} will give a most precise
determination of the mixing angle.

The attainable accuracy in the independent determination of the
relevant neutrino oscillation parameters from non-atmospheric neutrino
experiments~\cite{MINOS,T2K,FutureLBL} makes it possible to attempt an
inversion of the strategy: to use the oscillation parameters
(independently determined in non atmospheric neutrino experiments) as
inputs in the atmospheric neutrino analysis in order to extract the
atmospheric neutrino fluxes directly from the data.  One must notice,
however, that at present this independent determination of the
oscillation parameters is still incomplete, since neither MINOS nor
K2K have provided us with enough precision, in particular for what
concerns the relevant mixing angle.  Consequently such inversion of
the strategy is still not possible without further assumptions.  In
that respect our results can be regarded as the presentation of a novel
method to determine the atmospheric neutrino flux which, at present,
can only be applied under the assumption that oscillation parameters
will eventually be measured at LBL experiments and 
that they will have a
value close to the present best fit value.

There are several motivations for such direct determination of the
atmospheric neutrino fluxes.  First of all it would provide a
cross-check of the standard flux calculations as well as of the size
of the associated uncertainties (which, being mostly theoretical, are
difficult to quantify).  Second, a precise knowledge of the atmospheric
neutrino flux is of importance for high energy neutrino
telescopes~\cite{icecubelect}, both because they are the main
background and they are used for detector calibration.  Finally, such
program may quantitatively expand the physics potential of future
atmospheric neutrino experiments~\cite{hyperk,uno,ino}.  Technically,
however, this program is challenged by the absence of a generic
parametrization of the energy and angular functional dependence of the
fluxes which is valid in all the range of energies where there is
available data.

In this work we present the results of a first step in the direction
of this alternative approach on the determination of the atmospheric
neutrino fluxes: we will determine the energy dependence of the
atmospheric neutrino fluxes from the data on atmospheric neutrino
event rates measured by the Super-Kamiokande experiment. The present
experimental accuracy of the Super-Kamiokande experiment is not enough
to allow for a separate and precise determination of the energy,
zenith angle and flavour dependence of the atmospheric flux.
Consequently, in this work we will rely on the zenith and flavour
dependence of the flux as predicted by some of the atmospheric flux
calculations in Refs.~\cite{honda,bartol,others}, and discuss the
estimated uncertainty associated with this choice.  Furthermore, as
discussed above, the fluxes are determined under the assumption that
oscillation parameters will eventually be independently determined by
non atmospheric neutrino experiments with a value close to the present
best fit.

In our determination of the energy dependence of the atmospheric
neutrino fluxes, the problem of the unknown functional form for the
neutrino flux is bypassed by the use of neural networks as
interpolants.  Artificial neural networks have long been used in
different fields, from biology to high energy physics, and from
pattern recognition to business intelligence applications. In this
work we use artificial neural networks since they are a most unbiased
prior, that is, they allow us to parametrize the atmospheric neutrino
flux without having to assume any functional behavior.  Furthermore,
the determined flux comes together with a faithful estimation of the
associated uncertainties obtained using the Monte Carlo method for
error estimation.

Indeed the problem of the {\it deconvolution} of the atmospheric flux
from experimental data on event rates is rather close in spirit to the
determination of parton distribution functions in deep-inelastic
scattering from experimentally measured structure
functions~\cite{qcdbook,mcsrev}. For this reason, in this work we will
apply to the determination of the atmospheric neutrino fluxes a
general strategy originally designed to extract parton distributions
in an unbiased way with faithful estimation of the
uncertainties\footnote{This strategy has also been successfully
applied with different motivations in other contexts like tau lepton
decays~\cite{tau} and B meson
physics~\cite{bmeson}.}~\cite{f2ns,f2nnp,rojothesis,pdf2,heralhc,
nnpdfdis}.

The outline of this paper is as follows: in Section~\ref{sec:genstrat}
we describe our general strategy that we will use including the
description of the experimental data used, as well as the Monte Carlo
replica generation and the neural network training procedure. In
Section~\ref{sec:refit} we present the results for our reference fit,
and in Section~\ref{sec:stab} we analyze the impact of various choices
that define this reference. After concluding, four appendices
summarize the most technical aspects of this work: the definition of
several statistical estimators, an example of the Monte Carlo method
for error estimation, some details of the neural networks employed,
and a brief review of genetic algorithms.

\section{General Strategy}
\label{sec:genstrat}

The general strategy that will be used to determine the atmospheric
neutrino fluxes was first presented in Ref.~\cite{f2ns} and needs only
some modifications in order to be adapted to the problem under
consideration. The path from the data to the flux parametrization
involves two distinct stages~\cite{f2ns,f2nnp}. In the first stage, a
Monte Carlo sample of replicas of the experimental data on neutrino
event rates (``artificial data'') is generated. These can be viewed as
a sampling of the probability measure on the space of physical
observables at the discrete points where data exist. In the second
stage one uses neural networks to interpolate between these points. In
the present case, this second stage in turn consists of two sub-steps:
the determination of the atmospheric event rates from the atmospheric
flux in a fast and efficient way, and the comparison of the event
rates thus computed to the data in order to tune the best-fit form of
input neural flux distribution (``training of the neural network'').
Combining these two steps, the space of physical observables is mapped
onto the space of fluxes, so the experimental information on the
former can be interpolated by neural networks in the latter.

Let us now describe each stage in turn for the specific case discussed
in this paper.

\subsection{Experimental Data and Generation of Replicas}

In the present analysis we use the complete data on atmospheric
neutrino event rates from the phase-I of the Super-Kamiokande
experiment~\cite{sk}. Data from other atmospheric neutrino
experiments~\cite{macro, soudan2}, although important as a
confirmation, are not included in this work because of their lower
statistical significance. Higher energy data from neutrino telescopes
like Amanda are not publicly available in a format which allows for
its inclusion in the present analysis and its treatment is left for
future work.

The full Super-Kamiokande-I atmospheric neutrino data sample is
divided in 9 different types of events: contained events in three
energy ranges, Sub-GeV, Mid-GeV\footnote{In the Sub-GeV samples the
lepton momentum $p_l$ satisfies $p_l < 400$ MeV while in the Mid-GeV
samples the lepton momentum $p_l$ satisfies $p_l > 400$ MeV.} and
Multi-GeV electron- and muon-like, partially contained muon-like
events and upgoing stopping and thrugoing muon events. Each of the
above types of events is divided in 10 bins in the final state lepton
zenith angle $\phi_l$, with $-1\le \cos\phi_l \le 1$ for contained and
partially contained events and $-1\le \cos\phi_l \le 0$ for stopping
and thrugoing muon events. Therefore we have a total of $N_{\dat}=90$
experimental data points, which we label as
\begin{equation}
    R_i^{(\exp)}, \qquad i=1,\ldots,N_{\dat} \,.
\end{equation}

Note that each type of atmospheric neutrino event rate is sensitive to
a different region of the neutrino energy spectrum.  For example the
expected event rate for contained events can be computed as:
\begin{multline} \label{eq:contained}
    R_i = n_\text{tgt} T \sum_{\alpha,\beta,\pm}
    \int_0^\infty dh \int_{-1}^{+1} dc_\nu
    \int_{\Emin}^\infty dE_\nu \int_{\Emin}^{E_\nu} dE_l
    \int_{-1}^{+1} dc_a \int_0^{2\pi} d\varphi_a
    \frac{d^2 \Phi_\alpha^\pm}{dE_\nu \, dc_\nu}(E_\nu, c_\nu)
    \\
    \, \kappa_\alpha^\pm(E_\nu, c_\nu, h)
    \, P_{\alpha\to\beta}^\pm(E_\nu, c_\nu, h \,|\, \vec\eta)
    \, \frac{d^2\sigma_\beta^\pm}{dE_l \, dc_a}(E_\nu, E_l, c_a)
    \, \varepsilon_\beta^\text{bin}(E_l, c_l(c_\nu, c_a, \varphi_a))
    \,,
    \qquad
\end{multline}
where $P_{\alpha\to\beta}^+$ ($P_{\alpha\to\beta}^-$) is the
$\nu_\alpha \to \nu_\beta$ ($\bar{\nu}_\alpha \to \bar{\nu}_\beta$)
conversion probability for given values of the neutrino energy
$E_\nu$, the cosine $c_\nu$ of the angle between the incoming neutrino
and the vertical direction, the production altitude $h$, and the
oscillation parameters $\vec\eta$. Here $n_\text{tgt}$ is the number
of targets, $T$ is the experiment running time, $\Phi_\alpha^+$
($\Phi_\alpha^-$) is the flux of atmospheric neutrinos (antineutrinos)
of type $\alpha$, $\kappa_\alpha^\pm$ is the altitude distribution
(normalized to one) of the neutrino production point and
$\sigma_\beta^+$ ($\sigma_\beta^-$) is the charged-current neutrino-
(antineutrino-) nucleon interaction cross section.
The variable $E_l$ is the energy of the final lepton of type $\beta$,
while $c_a$ and $\varphi_a$ parametrize the opening angle between the
incoming neutrino and the final lepton directions as determined by the
kinematics of the neutrino interaction.
Finally, $\varepsilon_\beta^\text{bin}$ gives the probability that a
charged lepton of type $\beta$, energy $E_l$ and direction $c_l$
contributes to the given bin.

Correspondingly for upgoing muons the expected number of events in
each bin can be evaluated as:
\begin{multline} \label{eq:upgoing}
    R_i = \rho_\text{rock} T \sum_{\alpha,\pm}
    \int_0^\infty dh \int_{-1}^{+1} dc_\nu
    \int_{\Emin}^\infty dE_\nu
    \int_{\Emin}^{E_\nu} dE^0_\mu \int_{\Emin}^{E^0_\mu} d\Efin_\mu
    \int_{-1}^{+1} dc_a \int_0^{2\pi} d\varphi_a
    \\
    \frac{d^2 \Phi_\alpha^\pm}{dE_\nu \, dc_\nu}(E_\nu, c_\nu)
    \, \kappa_\alpha^\pm(E_\nu, c_\nu, h)
    \, P_{\alpha\to\mu}^\pm(E_\nu, c_\nu, h \,|\, \vec\eta)
    \, \frac{d^2\sigma_\mu^\pm}{dE^0_\mu \, dc_a}(E_\nu, dE^0_\mu, c_a)
    \\
    R_\text{rock}(E^0_\mu,\Efin_\mu)
    \, \mathcal{A}_\text{eff}^\text{bin}(\Efin_\mu,
    c_l(c_\nu, c_a, \varphi_a)) \,,
    \qquad
\end{multline}
where $\rho_\text{rock}$ is the density of targets in standard rock,
$R_\text{rock}$ is the effective muon range~\cite{Lipari:1991ut} for a
muon which is produced with energy $E^0_\mu$ and reaches the detector
with energy $\Efin_\mu$, and $\mathcal{A}_\text{eff}^\text{bin}$ is
the effective area for stopping and thrugoing muons respectively.

Thus in general
\begin{equation} \label{eq:weights}
    R_i= \int dE_\nu \frac{dR_i(E_\nu)}{dE_\nu} \,,
\end{equation}
where $\frac{dR_i(E_\nu)}{dE_\nu}$ gives the contribution of the
neutrino flux of energy $E_\nu$ to the event rate $R_i$ after
weighting with the neutrino interaction cross sections, muon
propagation and detection efficiencies. In order to illustrate which
part of the atmospheric neutrino spectrum is mostly determined by each
event type we plot in Fig.~\ref{fig:events} the function
$\frac{dR_i(E_\nu)}{dE_\nu}$ (normalized to one) for the Honda
atmospheric fluxes~\cite{honda}, for each event type -- Sub-GeV,
Mid-GeV and Multi-GeV, partially contained, stopping and thrugoing
muon events-- averaged over zenith angle (assuming oscillations with
$\Delta m_{\mathrm{atm}}^2=2.2\times 10^{-3}$ eV$^2$ and
$\tan^2\theta_{\mathrm{atm}}=1$). 
In particular from the figure we note that for energies larger than
$E_\nu\sim$ few TeV and smaller than $E_\nu\sim0.1$ GeV there is
essentially no information on the atmospheric neutrino flux coming
from the available event rates.

\FIGURE[t]{
  \includegraphics[width=0.75\textwidth]{fig-2a.epsi}
  \caption{\label{fig:events}%
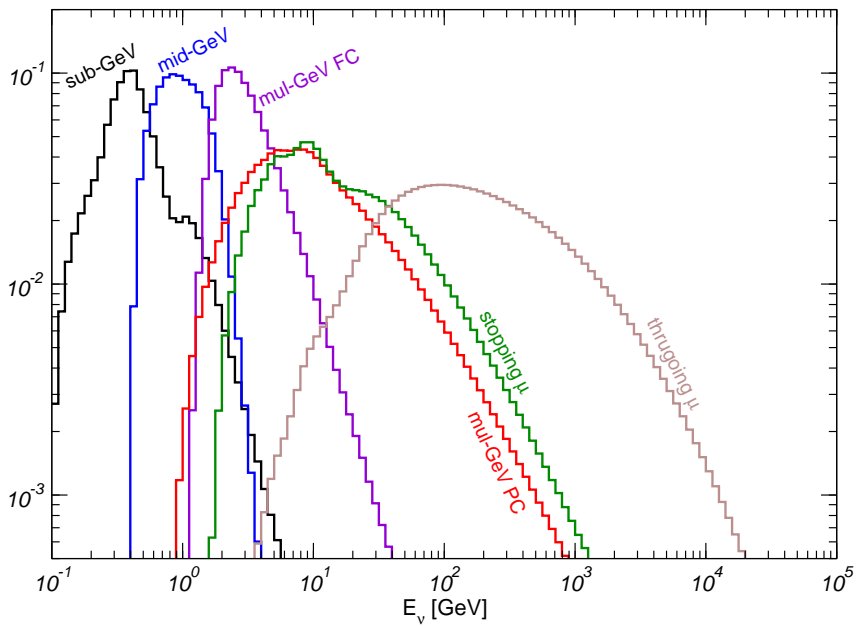
    Sensitivity to the neutrino energy of the different types of
    neutrino event rates (see Eq.~\eqref{eq:weights}). In this figure
    we assume oscillations with $\Delta m^2_{\mathrm{atm}}=2.2\times
    10^{-3}$ eV$^2$ and $\tan^2 \theta_{\mathrm{atm}}=1$ and the Honda
    atmospheric fluxes~\cite{honda}.}%
  }

The experimental correlation matrix for the 
event rates $R^{(\exp)}_i$ is
constructed in the following way:
\begin{equation} \label{eq:cormat}
    \rho^{(\exp)}_{ij}=\frac{\sigma^{\stat,2}_i\delta_{ij}
      + {\displaystyle \sum_{n=1}^{N_{\cor}}} \sigma^{\cor,n}_i 
      \sigma^{\cor,n}_j}{\sigma^{\tot}_{i}\sigma^{\tot}_{j}} \,,
\end{equation}
where the statistical uncertainty is given by
\begin{equation}
    \sigma^{\stat}_i=\sqrt{R_i^{(\exp)}} \,,
\end{equation}
and the $N_{\cor}$ correlated uncertainties are computed from the
couplings factors, $\pi_i^n$ to the corresponding {\it pull}
$\xi_{n}$~\cite{Fogli:2002pt},
\begin{equation}
    \sigma^{\cor,n}_i \equiv R_i^{(\exp)} \pi_i^n \,.
\end{equation}
The couplings $\pi_i^n$ used in the analysis are a generalization of
those given in Ref.~\cite{ournew} to include the separation between
sub-GeV and mid-GeV and the partially contained samples, and will be
described in more detail in a forthcoming publication.

In particular, we consider three different sets of correlated errors:
\begin{enumerate}
  \item Experimental systematic uncertainties ({\it exp}), like
    uncertainty in the detector calibration and efficiency.
    
  \item Theoretical cross-section uncertainties ({\it cross}), like
    cross-section normalization errors and cross section ratios
    errors.
    
  \item Theoretical zenith angle and flavor flux uncertainty ({\it
      flx}).
\end{enumerate}
Note that in standard neutrino oscillation parameter global fits
like~\cite{sk}, the flux uncertainties include on top of the above a
flux normalization and a flux tilt errors, but in our case we do not
have to include them because we are determining the energy dependence
of the atmospheric flux (including its normalization) directly from
the neutrino data. In other words, since our objective is to establish
how well we can determine the normalization and energy dependence of
the flux directly from the data, nothing is assumed about them:
neither their values nor their theoretical uncertainties.  We will go
back to this point after Eq.~\eqref{eq:fluxnn}.

Finally the total error is computed adding the statistical and
correlated errors in quadrature
\begin{equation} \label{eq:sigtot}
    \sigma^{\tot}_i=\sqrt{\sigma^{\stat,2}_i
      +\sum_{n=1}^{N_{\cor}} \lp \sigma^{\cor,n}_i\rp^2} \,.
\end{equation}
We summarize the characteristic values of the uncertainties in the
different data samples in Table~\ref{tab:datafeat}. From the table we
see that the experimental statistical errors are more important in the
high energy samples while the correlations are dominated by the cross
section uncertainties.


\TABLE[t]{
  \begin{tabular}{|c|
	@{\extracolsep{0.2cm}}c|
	@{\extracolsep{0.2cm}}c
	@{\extracolsep{0.2cm}}c
	@{\extracolsep{0.2cm}}c|
	@{\extracolsep{0.2cm}}c
	@{\extracolsep{0.2cm}}c
	@{\extracolsep{0.2cm}}c|
	@{\extracolsep{0.2cm}}c
	@{\extracolsep{0.2cm}}c
	@{\extracolsep{0.2cm}}c|}
      \hline
      &
      &\multicolumn{3}{c|}{\it exp}
      &\multicolumn{3}{c|}{\it exp+cross}
      &\multicolumn{3}{c|}{\it exp+cross+flx}\\
      \hline
      &$\la \sigma^{\stat}\ra$
      & $\la \sigma^{\mathrm{cor}}\ra$ &
      $\la \sigma^{\tot}\ra$ &
      $\la \rho\ra$
      & $\la \sigma^{\mathrm{cor}}\ra$ &
      $\la \sigma^{\tot}\ra$ &
      $\la \rho\ra$
      &$\la \sigma^{\mathrm{cor}}\ra$ &
      $\la \sigma^{\tot}\ra$ &
      $\la \rho\ra$ \\
      \hline
      {$\nu_e$ Sub-GeV}
      & 0.07 &  0.01 & 0.08 & 0.21
      & 0.15   & 0.17  & 0.83
      & 0.15   & 0.17  & 0.84
      \\\hline
      {$\nu_{\mu}$ Sub-GeV}
      & 0.09 & 0.01   & 0.09  & 0.20
      & 0.15   & 0.17  & 0.79
      & 0.15 & 0.18  & 0.79
      \\\hline
      {$\nu_e$ Mid-GeV}
      & 0.08  & 0.01  & 0.08   & 0.21
      & 0.10   & 0.13  & 0.69
      & 0.11 & 0.13  &0.69
      \\\hline
      {$\nu_{\mu}$ Mid-GeV}
      & 0.07  & 0.01   & 0.07  & 0.21
      & 0.11   & 0.13  & 0.74
      & 0.11  & 0.13  & 0.74
      \\\hline
      {$\nu_{e}$ Multi-GeV}
      & 0.12 &0.03   & 0.12 & 0.24
      & 0.09   & 0.15  & 0.50
      & 0.10 & 0.16 & 0.51
      \\\hline
      {$\nu_{\mu}$ Multi-GeV}
      &0.13  & 0.03   & 0.13  & 0.23
      &  0.10  & 0.16  & 0.49
      & 0.11 & 0.17  & 0.50
      \\\hline
      {$\nu_{\mu}$ PC}
      & 0.11 & 0.06  & 0.13 & 0.40
      & 0.12   & 0.16  &0.62
      & 0.12   & 0.16  & 0.68
      \\\hline
      {${\mu}$ Stop}
      & 0.16  & 0.07  &0.17  &0.30
      & 0.10   & 0.19  & 0.41
      & 0.10 & 0.19  & 0.41
      \\\hline
      {$\mu$ Thru}
      &0.08  & 0.02   & 0.08  & 0.23
      & 0.09  & 0.12 & 0.66
      & 0.10 & 0.13  & 0.67  \\
      \hline
      \hline
      Total
      & 0.10 & 0.03  & 0.11 &  0.03
      & 0.11  & 0.15 &  0.35
      & 0.12  & 0.16 &  0.36
      \\
      \hline
  \end{tabular}
  \caption{\label{tab:datafeat}%
    Features of experimental data atmospheric neutrino event rates
    from Super-Kamiokande~\cite{sk}. Experimental errors are given as
    percentages.}%
  }

The purpose of the artificial data generation is to produce a Monte
Carlo set of `pseudo--data', {\it i.e.}\ $N_{\rep}$ replicas of the
original set of $N_{\dat}$ data points:
\begin{equation} \label{eq:replicas}
    R^{(\art)(k)}_i; \qquad k=1,\dots,N_{\rep} \,,
    \quad i=1,\dots,N_{\dat} \,,
\end{equation}
such that the $N_{\rep}$ sets of $N_{\dat}$ points are distributed
according to an $N_{\dat}$--dimensional multi-gaussian distribution
around the original points, with expectation values equal to the
central experimental values, and error and covariance equal to the
corresponding experimental quantities.

This is achieved by defining
\begin{equation} \label{eq:gen}
    R_i^{(\art)(k)}=R_i^{(\exp)}+r_i^{(k)}\sigma_i^{\tot}, \qquad
    i=1,\ldots,N_{\dat} \,,\quad
    k=1,\ldots,N_{\rep} \,,
\end{equation}
where $N_{\rep}$ is the number of generated replicas of the
experimental data, and where $r_i^{(k)}$ are univariate gaussian
random numbers with the same correlation matrix as experimental data,
that is they satisfy
\begin{equation}
    \la r_i^{(k)}r_j^{(k)}\ra_{\rep}
    =\rho_{ij}^{(\exp)}+\mathcal{O}\lp\frac{1}{N_{\rep}}\rp \,.
\end{equation}
Because~\cite{cowan} the distribution of the experimental data
coincides (for a flat prior) with the probability distribution of the
value of the event rate $R_i$ at the points where it has been
measured, this Monte Carlo set gives a sampling of the probability
measure at those points.

Note that all errors considered (including correlated systematics and
theoretical uncertainties) must be treated as gaussian in this
framework. However, this does not imply that the resulting flux
probability density is gaussian, since the Monte Carlo method allows
for a non-gaussian distribution of best-fit atmospheric neutrino flux
distributions to be obtained as a result of the fitting procedure.

We can then generate arbitrarily many sets of pseudo--data, and choose
the number of sets $N_{\rep}$ in such a way that the properties of the
Monte Carlo sample reproduce those of the original data set to
arbitrary accuracy.  The relevant issue at this point is to determine
the minimum number of replicas required to reproduce the properties of
the original data set with enough accuracy.

In our particular case we have generated two different samples of
replicas, each with one different set of correlated uncertainties, as
described above, {\it exp+cross}, and {\it exp+cross+flx}. A fit
including the {\it exp} uncertainties only is meaningless since the
results of the Super-Kamiokande experiment are not consistent with the
neutrino oscillation hypothesis (since the overall $\chi^2_\text{SK}$
is too high) and with the results of K2K and MINOS (since the
preferred $\Delta m^2_\text{SK}$ is too low) if both flux and
cross-section uncertainties are simultaneously
neglected~\cite{rcctalk}.
The features of the Monte Carlo generated replicas with these two
types of uncertainties included in the generation are summarized in
terms of several statistical estimators (see
Appendix~\ref{sec:estimators} for definitions) in
Table~\ref{tab:gendata}.

These statistical estimators allow us to assess in a quantitative way
if the Monte Carlo sample of replicas reproduces the features of the
experimental data.  For example, we can check that averages, variance
and covariance of the pseudo-data reproduce central values and
covariance matrix elements of the original data.  From the table we
see that for the two sets of uncertainties considered, 100 replicas
are enough to reproduce the properties of the original data set with
the required accuracy, namely a few percent, which is the accuracy of
the experimental data while it is clear that 10 replicas fall short to
reproduce the statistical properties of the experimental data.


\TABLE[t]{
  \hspace{5mm}
  \begin{tabular}{|c|ccc|}
      \hline
      &  {\it exp+cross}
      &  \multicolumn{2}{c|}{\it exp+cross+flx} \\
      $N_{\rep}$ & $\quad$100$\quad$ & $\quad$100$\quad$ &
      $\quad$ 10 $\quad$ \\
      \hline
      $\la PE\lc\la R \ra_{\rep}\rc\ra$  & 1.5\% & 1.5\%  & 4.7\% \\
      $r \lc R\rc $ & 0.99 & 0.99 & 0.99  \\
      \hline
      $\la PE\lc \sigma^{(\art)}\rc\ra_{\dat}$ & 10.5\% & 9.6\% & 10.1\% \\
      $\la \sigma^{(\exp)}\ra_{\dat}$ & 18.1  & 18.5 & 18.5 \\
      $\la \sigma^{(\art)}\ra_{\dat}$ & 17.5  & 18.9 & 13.8  \\
      $r\lc \sigma^{(\art)}\rc$       & 0.99 & 0.99  & 0.76 \\
      \hline
      $\la \rho^{(\exp)}\ra_{\dat}$ & 0.35 & 0.36  & 0.36 \\
      $\la \rho^{(\art)}\ra_{\dat}$ & 0.34 & 0.34  & 0.26  \\
      $r\lc \rho^{(\art)}\rc$       & 0.94  & 0.94  & 0.36  \\
      \hline
      $\la \mathrm{cov}^{(\exp)}\ra_{\dat}$   & 137.6  & 145.7  & 145.7 \\
      $\la \mathrm{cov}^{(\art)}\ra_{\dat}$   & 122.7  & 129.2   & 53.2 \\
      $r\lc \mathrm{cov}^{(\art)}\rc$         & 0.98 & 0.98    & 0.42   \\
      \hline
  \end{tabular}
  \hspace{5mm}
  \caption{\label{tab:gendata}%
    Comparison between experimental data and Monte Carlo data for
    different types of uncertainties and different numbers of
    $N_{\rep}$ included in the pseudo-data generation.}%
  }


\subsection{Neural Network Training}
\label{sec:training}

The second step consists of training $N_{\rep}$ neural networks.  In
our case each neural network parametrizes a differential flux, which
in principle should depend on the neutrino energy $E_\nu$, the zenith
angle $\cos \lp \theta_{\nu}\rp\equiv c_\nu$ and the neutrino {\it
type} $t$ ($t=1,\ldots,4$ labels the the neutrino flavor: electron
neutrinos and antineutrinos, and muon neutrinos and antineutrinos),
and is based on all the data in one single replica of the original
data set. 
However, the precision of the available experimental data is not
enough to allow for a separate determination of the energy, zenith
angle and type dependence of the atmospheric flux. Consequently in
this work we will assume the zenith and type dependence of the flux to
be known with some precision and extract from the data only its energy
dependence. Thus the neural flux parametrization will be:
\begin{equation} \label{eq:fluxnn}
    \Phi^{(\net)}
    \lp E_{\nu},c_{\nu},t\rp \equiv 
    \frac{d^2\Phi^{(\net)}_t}{dE_\nu dc_\nu}
    =\NNe 
    \frac{d^2\Phi^{\mathrm{(ref)}}_t}{dE_\nu\, dc_\nu}
\end{equation}
where $\NNe$ is the neural network output when the input is the
neutrino energy $E_\nu$ 
\begin{equation}
    \NNe\equiv\mathrm{NN}\lp E_\nu,\vec\omega \rp  \,.
\end{equation}
and it depends on the neutrino energy\footnote{In fact the input
variable is rather $\log_{10} E_\nu$, since it can be
shown~\cite{nnthesis} that an appropriate {\it preprocessing} of the
input parameters speeds up the training process.} as well as on the
parameters $\vec\omega$ of the neural network (see Appendix
\ref{sec:neuralnetdet} for details).  In Eq.~\eqref{eq:fluxnn}
$\Phi^{(\mathrm{ref})}$ is a reference differential flux, which we
take to be the most recent computations of either the
Honda~\cite{honda} or the Bartol~\cite{bartol} collaborations,
extended to cover also the high-energy region by consistent matching
with the Volkova fluxes~\cite{volkova}. Notice that in what respects
the normalization and energy dependence of the fluxes, the choice of
reference flux is irrelevant.  Any variation on the normalization or
on the energy dependence of the reference flux can be compensated by
the corresponding variation of $\NNe$ so that the output flux
$\Phi^{(\net)}$ will be the same. The dependence of the results of the
analysis on the reference flux comes because of the differences among
the different flux calculations in angular and flavour dependence.

Nothing further is assumed about the function $\NNe$ whose value is
only known after the full procedure of training of the neural net 
is finished.  There
are, however, some requirements about the choice of the architecture
of the neural network.  As discussed in Refs.~\cite{f2ns,rojothesis}
such choice cannot be derived from general rules and it must be
tailored to each specific problem. The main requirements for an
optimal architecture are first of all that the net is large enough so
that the results are stable with respect small variations of the
number of neurons (in this case the neural net is called {\it
redundant}) and, second, that this net is not so large than the
training times become prohibitive. For our problem the neural network
must have a single input neuron (whose value is $\log(E_\nu$)) and a
final output neuron (whose value is the $\NNe$) and a number of hidden
layers with several neurons each (see appendix \ref{sec:neuralnetdet}
for further details).  We have checked that an architecture with two
hidden layers with 5 neurons each, {\it i.e.}\ a 1-5-5-1 network,
satisfies the above requirements in the present case.

The process which determines the function $\NNe$ which better
describes each of the $k=1,\dots,N_{\rep}$ sets of artificial data,
$\{R_i^{(\art)(k)}\}$, is what we call training of the neural network.
It involves two substeps. First for a given $\NNe$ the expected
atmospheric event rates have to be computed in a fast an efficient
way. Second the neural network parameters $\vec\omega$ have to be
determined by minimizing some error function. We describe them next.

\subsubsection{From Atmospheric Fluxes to Event Rates}
\label{sec:rates}

The expected event rate for contained and upgoing muon events for a
given set of neural network parameters $\vec\omega$, or what is the
same for a given value of the neural network flux, can be obtained by
substituting Eq.~\eqref{eq:fluxnn} into Eqs.~\eqref{eq:contained} and
\eqref{eq:upgoing} respectively. However, from a practical point of
view the above expressions are very time-consuming to evaluate, and
this is a very serious problem in our case since in the neural network
approach one requires a very large number of evaluations in the
training process. 

The procedure can be speed up if one realizes that for a given flux
$\Phi^{(\net)}$ the expected event rates can always be written as
\begin{equation} \label{eq:rinet}
    R_i^{(\net)} = \sum_t \int_{-1}^1 dc_\nu
    \int_{E_{\mathrm{min}}}^{\infty}
    dE_\nu \, \Phi^{(\net)}(E_\nu,c_\nu,t)
    \, \widetilde{C}_i(E_\nu,c_\nu,t) \,,
\end{equation}
where $t \equiv \{\alpha,\pm\}$ labels both the flavor and the
chirality of the initial neutrino state. Comparing
Eq.~\eqref{eq:rinet} with Eqs.~\eqref{eq:contained} and
\eqref{eq:upgoing} we get:
\begin{multline} \label{eq:Ccontanined}
    \widetilde{C}_i(E_\nu,c_\nu,t)=
    n_\text{tgt} T \sum_{\beta}
    \int_0^\infty dh \int_{\Emin}^{E_\nu} dE_l
    \int_{-1}^{+1} dc_a \int_0^{2\pi} d\varphi_a
    \\
    \, \kappa_t(E_\nu, c_\nu, h)
    \, P_{t \to\beta}(E_\nu, c_\nu, h \,|\, \vec\eta)
    \, \frac{d^2\sigma_\beta}{dE_l \, dc_a}(E_\nu, E_l, c_a)
    \, \varepsilon_\beta^\text{bin}(E_l, c_l(c_\nu, c_a, \varphi_a))
    \qquad
\end{multline}
for contained events, and
\begin{multline} \label{eq:Cupgoing}
    \widetilde{C}_i(E_\nu,c_\nu,t) = \rho_\text{rock} T
    \int_0^\infty dh 
    \int_{\Emin}^{E_\nu} dE^0_\mu \int_{\Emin}^{E^0_\mu} d\Efin_\mu
    \int_{-1}^{+1} dc_a \int_0^{2\pi} d\varphi_a
    \\
    \, \kappa_t(E_\nu, c_\nu, h)
    \, P_{t \to\mu}(E_\nu, c_\nu, h \,|\, \vec\eta)
    \, \frac{d^2\sigma_\mu}{dE^0_\mu \, dc_a}(E_\nu, dE^0_\mu, c_a)
    \\
    R_\text{rock}(E^0_\mu,\Efin_\mu)
    \, \mathcal{A}_\text{eff}^\text{bin}(\Efin_\mu,
    c_l(c_\nu, c_a, \varphi_a))
    \qquad\quad
\end{multline}
for upgoing-muon events. Consequently, if one discretizes the
differential fluxes in $N_e$ energy intervals $I_e$ and $N_z$ zenith
angle intervals $I_z$ one can write
\begin{gather}
    \label{eq:decom}
    \Phi^{(\net)}(E_\nu,c_\nu,t) \simeq
    \sum_{e,z} \Psi_{ezt}^{(\net)}
    \, \theta(E_\nu\in I_e)
    \, \theta(c_\nu \in I_z) \,,
    \\
    e=1,\ldots,N_e, \qquad z=1,\ldots,N_z  \,, \qquad t=1,\ldots,4 \,,
\end{gather}
and therefore it is possible to write the theoretical predictions as a
sum of the elements of the discretized flux table,
\begin{equation} \label{eq:coef}
    R_i^{(\net)}=\sum_{ezt}C^i_{ezt}\Psi^{(\net)}_{ezt} \,,
\end{equation}
where the coefficients $C^i_{ezt}$, which are the most time-consuming
ingredient, need only to be precomputed once before the training,
since they do not depend on the parametrization of the atmospheric
neutrino flux.

In our calculations we have used $N_e=100$ energy intervals, equally
spaced between $\log_{10} E_\nu=-1$ and $\log_{10} E_\nu=5$ (with
$E_\nu$ in GeV) and $N_{z}=80$ bins in zenith angle, equally spaced
between $c_\nu=-1$ and $c_\nu=1$ so
\begin{equation} \label{eq:ftab}
    \Psi_{ezt}^{(\net)}\equiv\int_{-1+6(e-1)/N_e}^{-1+6e/N_e} d \log_{10} E_\nu ~
    \int_{-1+(z-1)/N_z}^{-1+z/N_z} dc_\nu ~~\Phi^{(\net)}(E_\nu,c_\nu,t)\,,
\end{equation}
where the integrations in Eq.~\eqref{eq:ftab} are performed via Monte
Carlo numerical integration.

\subsubsection{Minimization Procedure}

From each replica of artificial data $\{R_i^{(\art)(k)}\}$ an
atmospheric neutrino flux parametrized with a neural network
$\Phi^{(\net)(k)}$ is obtained. The $N_{\dat}$ data points in each
replica are used to determine the parameters of the associated neural
net. The fit of the fluxes to each replica of the data, or what is the
same, the determination of the parameters that define the neural
network, its weights, is performed by maximum likelihood. This
procedure, the so-called neural network training, proceeds by
minimizing an error function $E^{(k)}$, which coincides with the
$\chi^2$ of the experimental points when compared to their theoretical
determination obtained using the given set of fluxes:
\begin{equation} \label{eq:chi2}
    E^{(k)}\lp \vec\omega\rp = \mathrm{min}_{\vec\xi}
    \lc \sum_{i=1}^{N_{\dat}} \lp \frac{R_i^{\mathrm{(\net)(k)}}\lp
      \vec\omega\rp
      \lc 1+{\displaystyle \sum_n} \pi_i^n\xi_n\rc-R_i^{(\art) (k)}}{
      \sigma_i^{\stat}}\rp^2+\sum_n\xi^2_n \rc \,,
\end{equation}
The case $k=0$ corresponds to the experimental values,
$R_i^{(\art)(0)}=R_i^{(\exp)}$. The $E^{(k)}$ has to be minimized with
respect to $\vec\omega$, the parameters of the neural network.

We perform two different type of fits which we denote by {\it
exp+cross} and {\it exp+cross+flx}. To be consistent we include in
each one the same correlated uncertainties that have been included in
the replica generation. For example, if we want to include only the
effects of the {\it exp+cross} uncertainties, in
Eqs.~\eqref{eq:cormat} and \eqref{eq:sigtot} the sum includes only
experimental systematic and theoretical cross section uncertainties,
while in Eq.~\eqref{eq:chi2} one imposes $\xi_i=0$ for {\it flx}
uncertainties.

Unlike in conventional fits with errors, however, the covariance
matrices of the best--fit parameters are irrelevant and need not be
computed. The uncertainty on the final result is found from the
variance of the Monte Carlo sample. This eliminates the problem of
choosing the value of $\Delta\chi^2$ which corresponds to a one-sigma
contour in the space of parameters.

Rather, one only has to make sure that each neural net provides a
consistent fit to its corresponding replica. If the underlying data
are incompatible or have underestimated errors, the best fit might be
worse than one would expect with properly estimated gaussian errors
--- for instance in the presence of underestimated errors it will have
typically a value of $\chi^2$ per degree of freedom larger than one.
However, neural nets are ideally suited for providing a fit in this
situation, based on the reasonable assumption of smoothness: for
example, incompatible data or data with underestimated errors will
naturally be fitted less accurately by the neural net. Also, this
allows for non-gaussian behavior of experimental uncertainties.

The minimization of Eq.~\eqref{eq:chi2} is performed with the use of
genetic algorithms (summarized in 
Appendix D, see Ref.~\cite{rojothesis} and references
therein for a more complete description). 
Because of the nonlinear dependence of the neural net on its
parameters, and the nonlocal dependence of the measured quantities on
the neural net (event rates are given by multi-dimensional
convolutions of the initial flux distributions), a genetic algorithm
turns out to be the most efficient minimization method. The use of a
genetic algorithm is particularly convenient when seeking a minimum in
a very wide space with potentially many local minima, because the
method handles a population of solutions rather than traversing a path
in the space of solutions.

The minimization is ended after a number of iterations of the
minimization algorithm large enough so that $E^{(k)}$ of 
Eq.~\eqref{eq:chi2} stops decreasing, that is, when the fit has
converged.\footnote{Note that the standard criterion to stop neural
network training, the overlearning criterion, cannot be used in our
case due to the scarce amount of data.} Thus an important issue in the
procedure is to determine the right number of iterations which should
be used.  In order to determine them, we define the total
$\chi^2_{\tot}$,
\begin{equation} \label{eq:chi2tot}
    \chi^{2}_{\tot} \equiv\mathrm{min}_{\vec\xi}
    \lc \sum_{i=1}^{N_{\dat}} \lp \frac{\la R_i^{\mathrm{(\net)}}\ra_{\rep}
      \lc 1+{\displaystyle \sum_n} \pi_i^n\xi_n\rc-R_i^{(\exp)}}{
      \sigma_i^{\stat}}\rp^2+\sum_n\xi^2_n \rc \,,
\end{equation}
with the event rates computed as an average over the sample of trained
neural nets,
\begin{equation} \label{eq:Rnet}
    \la R^{(\net)}_i\ra_{\rep}=\frac{1}{N_{\rep}} \sum_{k=1}^{N_{\rep}}
    R^{(\net)(k)}_i \,,
\end{equation}
and we study its value as a function of the number of iterations used
in the minimization.

We show in Fig.~\ref{fig:chi2gen} the dependence of $\chi^2_{\tot}$ on
the number of minimization iterations, in the two cases considered.
From the figure we see that the number of iterations needed to achieve
convergence can be safely taken to be $N_{it}=150$ both for the in the
{\it exp+cross} and {\it exp+cross+flx} fits.

\FIGURE[t]{
  \includegraphics[width=0.75\textwidth]{chi2gen.epsi}
  \caption{\label{fig:chi2gen}%
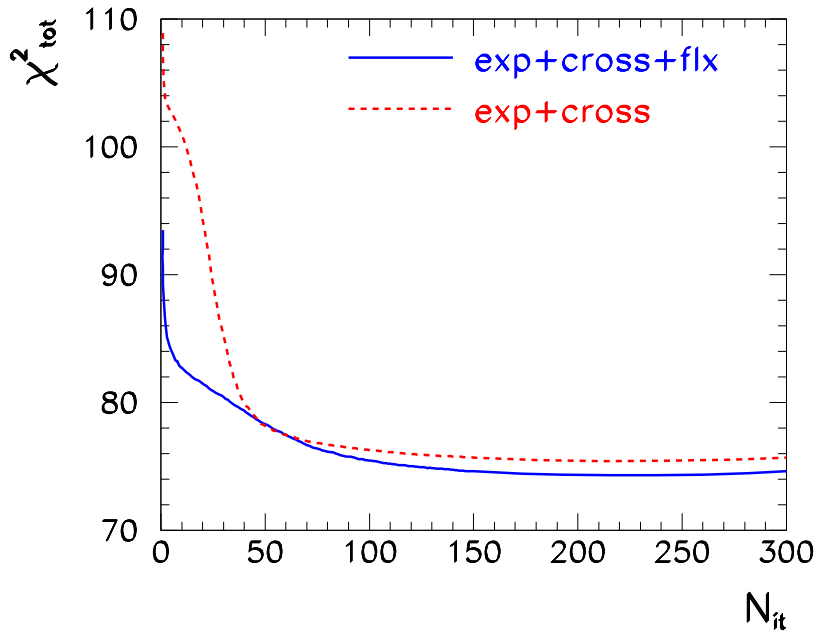
    Dependence of the $\chi^2_{\tot}$ with the number of iterations in
    the minimization procedure, for fits with different sets of
    uncertainties incorporated.}%
  }

Thus at the end of the procedure, we end up with $N_{\rep}$ fluxes,
with each flux $\Phi^{(\net)(k)}$ given by a neural net. The set of
$N_{\rep}$ fluxes provide our best representation of the corresponding
probability density in the space of atmospheric neutrino fluxes: for
example, the mean value of the flux at a given value of $E_\nu$ is
found by averaging over the replicas, and the uncertainty on this
value is the variance of the values given by the replicas.  Generally,
we expect the uncertainty on our final result to be somewhat smaller
than the uncertainty on the input data, because the information
contained in several data points is combined.

There are two type of tests that can be performed on the properties of
this probability measure.  First, the self-consistency of the Monte
Carlo sample can be tested in order to ascertain that it leads to
consistent estimates of the uncertainty on the final set of fluxes,
for example by verifying that the value of the flux extracted from
different replicas indeed behaves as a random variable with the stated
variance. This set of tests allows us to make sure that the Monte
Carlo sample of neural nets provides a faithful and consistent
representation of the information contained in the data on the
probability measure in the space of fluxes, and in particular that the
value of fluxes and their (correlated) uncertainties are correctly
estimated.

Furthermore the properties of this measure can be tested against the
input data by using it to compute means, variance and covariances
which can be compared to the input experimental ones which have been
used in the flux determination.

\section{Results for the Reference Fit}
\label{sec:refit}

In this section we discuss our results for the reference fit of the
atmospheric neutrino fluxes.  For this fit we use the
Honda~\cite{honda} flux as reference, the {\it exp+cross+flx} set of
uncertainties. and we assume $\nu_\mu\rightarrow \nu_\tau$
oscillations with oscillation parameters $\Delta m^2_{\mathrm{atm}}
=2.2\times 10^{-3}$ and $\tan^2\theta_{\mathrm{atm}}=1$.  We will
discuss in next section the dependence of the results on these
choices.

As described in the previous section, we start by generating a set of
$N_{\rep}=100$ replicas of the experimental data points according to
Eq.~\eqref{eq:gen} where in $\sigma_i^{\tot}$ we include both the
statistical as well as the correlated errors from experimental
systematic uncertainties, theoretical cross section uncertainties and
the theoretical flux uncertainties in the angular distributions and
neutrino type ratios. As shown in Table~\ref{tab:gendata}, this Monte
Carlo sample of replicas reproduces with enough precision the
statistical features of the original experimental data.

After that we proceed to the training of the neural networks as
described in Sec.~\ref{sec:training}.  Once the training of the sample
of neural networks has been completed, we obtain the set of $N_{\rep}$
$\Phi^{(\net)(k)}$ fluxes which provide us with the probability
density in the space of atmospheric neutrino fluxes. In particular we
compute the average atmospheric neutrino flux as
\begin{equation} \begin{split}
    \la \Phi^{(\net)}\ra_{\rep}(E_\nu,c_\nu,t) 
    &= \frac{1}{N_{\rep}} \sum_{k=1}^{N_{\rep}}
    \Phi^{(\net)(k)}(E_\nu,\cz,t)
    \\
    &= \lc \frac{1}{N_{\rep}} \sum_{k=1}^{N_{\rep}}
    NN^{(k)}(E_\nu)\rc \Phi^{(\mathrm{ref})}(E_\nu,\cz,t) \,,
\end{split} \end{equation}
and the standard deviation as
\begin{equation}
    \sigma_{\Phi}^2(E_\nu,c_\nu,t) = 
    \frac{1}{N_{\rep}} \sum_{k=1}^{N_{\rep}}
    \lp \Phi^{(\net)(k)}(E_\nu,c_\nu,t) \rp^2
    - \la \Phi^{(\net)}\ra_{\rep}^2 (E_\nu,c_\nu,t) \,.
\end{equation}
for any given value of the energy $E_\nu$, the zenith angle $\cz$ and
the neutrino type $t$.

In Fig.~\ref{fig:fluxref} we show the results for the flux (in
particular we show the angular averaged muon neutrino flux) as
compared with the computations of the Honda~\cite{honda} and
Bartol~\cite{bartol} groups. The results of the neural network fit are
shown as the $\la \Phi^{(\net)}\ra_{\rep}\pm \sigma_{\Phi}$ band as a
function of the neutrino energy. We see from the figure that the flux
obtained from this fits is in reasonable agreement with the results
from the the calculations of Honda and Bartol groups. We also see that
at lower energies the present uncertainty in the extracted fluxes is
larger than the range of variations between calculations while at
higher energies the opposite holds. The fit also seems to prefer a
slightly higher flux at higher energies.

\FIGURE[t]{
  \includegraphics[width=0.70\textwidth]{fluxref.epsi}
  \caption{\label{fig:fluxref}%
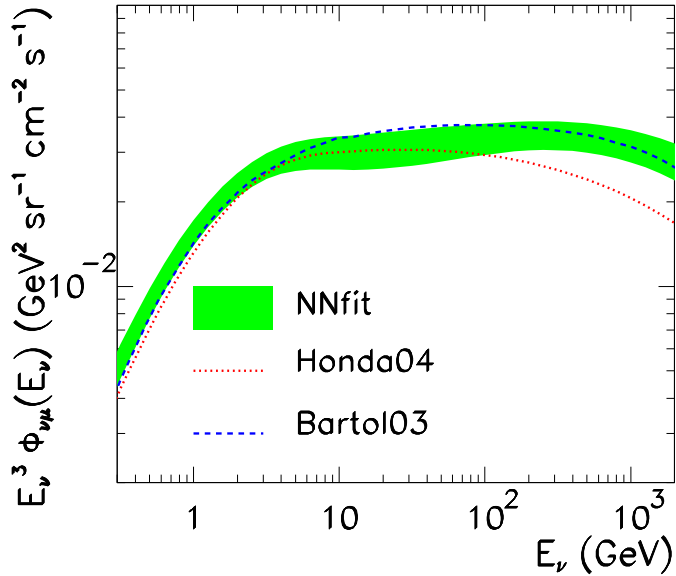
    Results for the reference fit for the angular averaged muon
    neutrino flux and comparison with numerical computations.}%
  }

The statistical estimators for this reference training are given in
the first column in Table~\ref{tab:resest}, where the different
estimators can be found in Appendix A. Note that errors are somewhat
reduced, as expected if the neural network has found the underlying
physical law, correlations increase and covariances are appropriately
reproduced, a sign that the sample of trained neural networks
correctly reproduces the probability measure that underlies
experimental data. We will return to the discussion on error reduction
in next section.


\TABLE[t]{
  \begin{tabular}{|c|c|c|c|}
      \hline
      Fit & Reference  & Bartol Flux & {\it exp+cross} \\
      \hline
      $\chi^2_{\tot}$  &    74.6 & 73.3 & 75.7
      \\
      \hline
      $\qquad \qquad \la PE\lc\la R \ra_{\rep}\rc\ra \qquad
      \qquad$ & $\qquad$  8.9\% $\qquad$
      & $\qquad$  9.1\% $\qquad$  & $\qquad$   9.0\% $\qquad$
      \\
      $r \lc R\rc $ & 0.99   & 0.98  &   0.99
      \\
      \hline
      $\la \sigma^{(\exp)}\ra_{\dat}$ & 18.5   & 18.5 & 18.1
      \\
      $\la \sigma^{(\net)}\ra_{\dat}$ & 14.7  &14.7  & 3.2 \\
      $r\lc \sigma^{(\net)}\rc$ & 0.97 & 0.98 & 0.81 \\
      \hline
      $\la \rho^{(\exp)}\ra_{\dat}$ & 0.36  & 0.36 & 0.35  \\
      $\la \rho^{(\net)}\ra_{\dat}$ & 0.65  & 0.66 & 0.78 \\
      $r\lc \rho^{(\net)}\rc$       & 0.69  & 0.71 & 0.74 \\
      \hline
      $\la \mathrm{cov}^{(\exp)}\ra_{\dat}$ & 145.7  &145.7 & 137.6 \\
      $\la \mathrm{cov}^{(\net)}\ra_{\dat}$ & 136.0  &131.8 & 8,5 \\
      $r\lc \mathrm{cov}^{(\net)}\rc$       & 0.95 &0.97 &  0.78\\
      \hline
  \end{tabular}
  \caption{\label{tab:resest}%
    Comparison between experimental data and the results of the neural
    network training for the reference fit (Honda fluxes as reference
    with {\it exp+cross+flx} errors), for the fit with Bartol fluxes
    as reference with {\it exp+cross+flx} errors and for Honda
    reference fluxes with with {\it exp+cross} errors only.  In all
    cases $N_{\rep}=100$ replicas are used.}%
  }

In Fig.~\ref{fig:rel_err}, we plot the relative error
$\sigma_\Phi/\Phi$ as function of the energy which, as seen in the
figure, grows at the lowest and highest energies. The fact that the
relative error grows in the region where less data is available
reflects the fact that the behavior of neural networks in those
regions is not determined by its behavior in the regions where more
data is available, as it would happen in fits with usual functional
forms.

\FIGURE[t]{
  \includegraphics[width=0.65\textwidth]{errel.epsi}
  \caption{\label{fig:rel_err}%
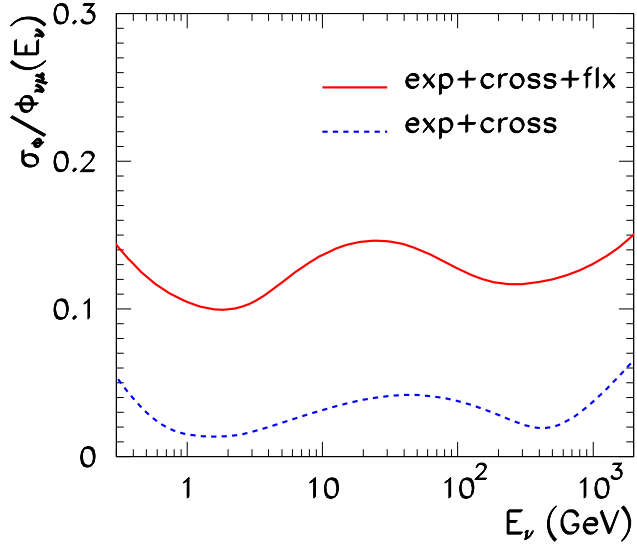
    Relative error in the determination of the flux.}%
  }

The first and second derivatives of the flux ratio with the associated
uncertainties,
\begin{align}
    \label{eq:derf1}
    D_1 \Phi^{(\net)}(E_\nu) &\equiv
    \frac{d}{d\ln E_\nu}
    \frac{\Phi^{(\net)}(E_\nu)}{\Phi^{(\mathrm{ref})}(E_\nu)} \,,
    \\[1mm]
    \label{eq:derf2}
    D_2 \Phi^{(\net)}(E_\nu) &\equiv
    \frac{d^2}{d^2\ln E_\nu}
    \frac{\Phi^{(\net)}(E_\nu)}{\Phi^{\mathrm{(ref)}}(E_\nu)} \,,
\end{align}
are shown in Fig.~\ref{fig:fluxder}. From the figure we see that,
within the present errors, the first derivative is not a constant, or
in other words one cannot parametrize the energy dependence of the
flux uncertainty as a simple {\it tilt} correction,
\begin{equation}
    \Phi^{(\net)}(E_\nu)=
    \Phi^{(\mathrm{ref})}(E_\nu)\lp 1+\delta \ln E_\nu\rp \,.
\end{equation}
However, we also see that the second derivative is compatible with
zero within errors in almost all the energy range, which implies that
the uncertainty is not a much more strongly varying function of the
energy.

\FIGURE[t]{
  \includegraphics[width=0.70\textwidth]{fluxer.epsi}
  \caption{\label{fig:fluxder}%
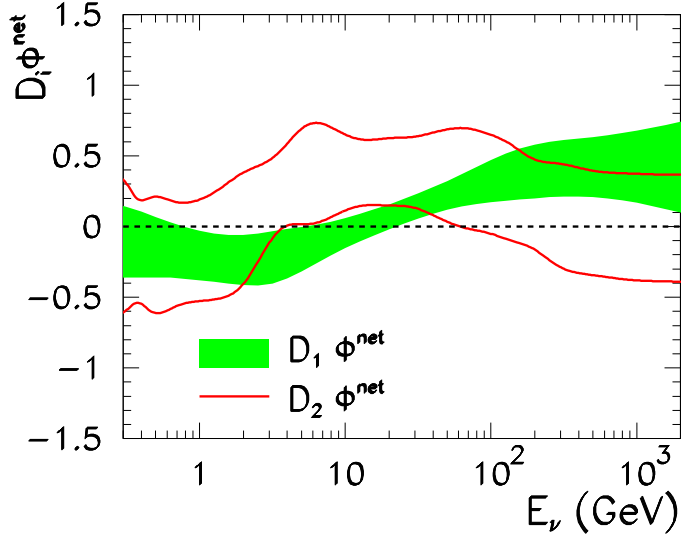
    First and second derivative of the atmospheric neutrino flux
    ratio, Eqs.~\eqref{eq:derf1} and \eqref{eq:derf2}.}%
  }

Finally from Eq.~\eqref{eq:coef} we compute the {\it predicted} event
rates for the $N_{\rep}$ fluxes $R^{(\net)(k)}_i$ and define their
average as in Eq.~\eqref{eq:Rnet} and their standard deviation as
\begin{equation} \label{eq:Rnetsig}
    \sigma_{R_i}^2=
    \frac{1}{N_{\rep}} \sum_{k=1}^{N_{\rep}}
    \lp R^{(\net)(k)}_i\rp^2 - \la R^{(\net)}_i\ra_{\rep}^2\,.
\end{equation}
In Fig.~\ref{fig:data} we show a comparison of the experimental data
with the corresponding predictions of the neural network
parametrization of the atmospheric neutrino flux. From the figure we
see that the predicted rates are in good agreement with the data, but,
as expected, have a smoother zenith angular dependence.

\FIGURE[t]{
  \includegraphics[width=\textwidth]{events_ref.epsi}
  \caption{\label{fig:data}%
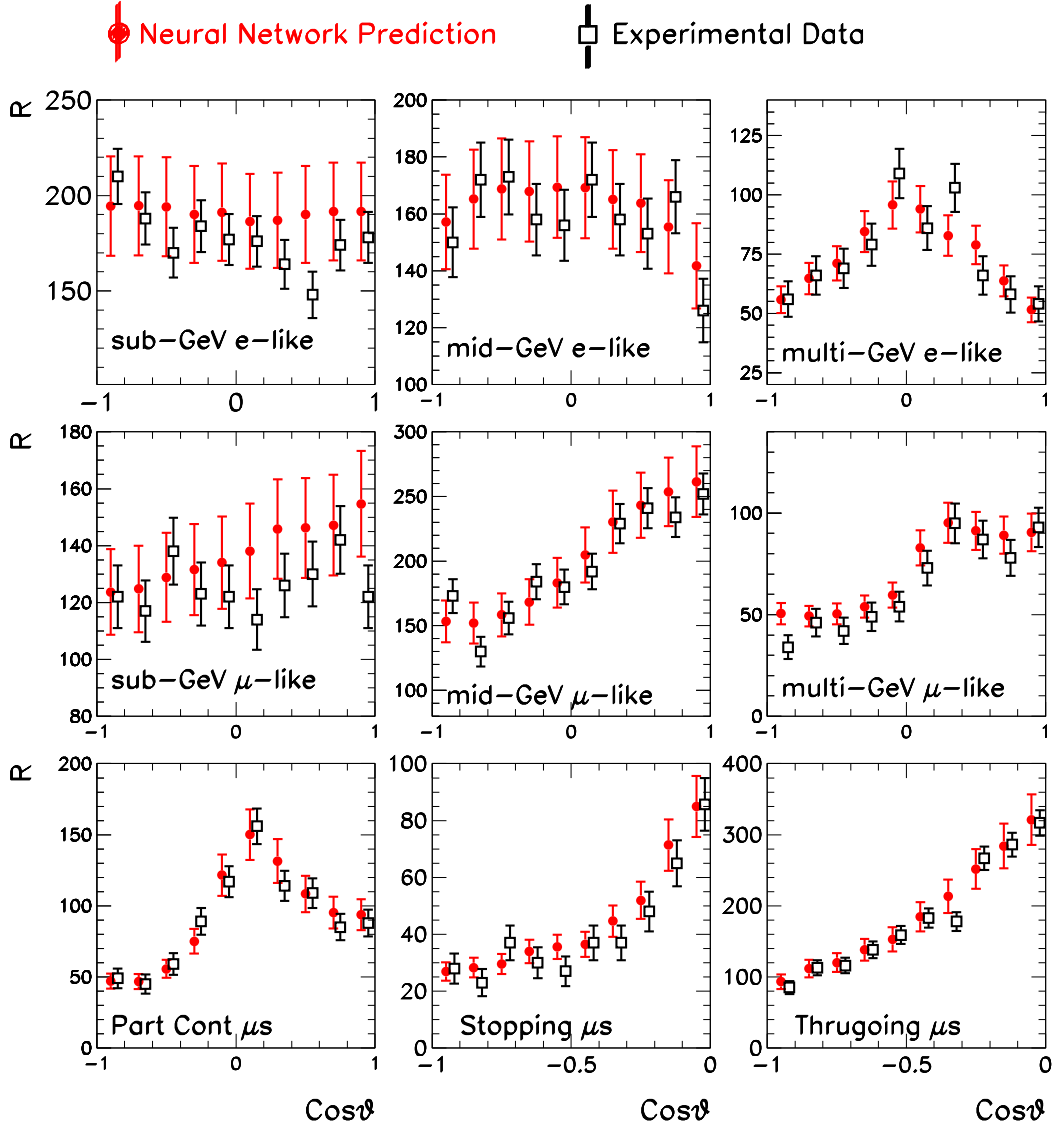
    Predicted number of atmospheric neutrino events using the
    atmospheric fluxes resulting from the reference fit compared to
    the experimental data points.  The central values correspond to
    the average prediction Eq.~\eqref{eq:Rnet} and the error bars give
    the 1$\sigma$ ranges Eq.~\eqref{eq:Rnetsig}. Notice that only the
    statistical error is shown for the experimental data points.}%
  }

Another interesting figure of merit to verify the correct statistical
behavior of the neural network training is the distribution of
$\chi^{2(k)}_{\dat}$, defined as the $\chi^2$ of the set of neural
fluxes compared to experimental data, that is
\begin{equation} \label{eq:chi2dat}
    \chi^{2(k)}_{\dat} =\mathrm{min}_{\vec\xi}
    \lc {\displaystyle \sum_{i=1}^{N_{\dat}}} \lp \frac{R_i^{\mathrm{(\net)(k)}}
      \lc 1+{\displaystyle \sum_n} \pi_i^n\xi_n\rc-R_i^{(\exp) }}{
      \sigma_i^{\stat}}\rp^2+\sum_n\xi^2_n \rc \,.
\end{equation}
This distribution is shown in Fig.~\ref{fig:chi2dist}. Note that for
all the fluxes $\chi^{2(k)}_{\dat} \ge \chi^2_{\tot}$. Furthermore, if
we define $\Delta\chi^2=\chi^{2(k)}_{\mathrm{max}} -\chi^2_{\tot}$,
where $\chi^{2(k)}_{\mathrm{max}}$ is the maximum value of the set of
68\% fluxes with lower $\chi^{2(k)}_{\dat}$,we get that in the present
analysis the 1-$\sigma$ range of fluxes obtained from the sample of
trained neural networks corresponds to $\Delta\chi^2\sim 5$, which
satisfies $\Delta\chi^2 \le \sqrt{2N_{\dat}}\sim 13$ as expected for a
consistent distribution of fits.

\FIGURE[t]{
  \includegraphics[width=0.63\textwidth]{chi2dist.epsi}
  \caption{\label{fig:chi2dist}%
    Distribution of $\chi^{2(k)}_{\dat}$, Eq.~\eqref{eq:chi2dat}, in
    the reference case.}%
  }

\section{Stability of the Results}
\label{sec:stab}

In this section we discuss the stability of our results with respect
to different inputs used for the reference fit, such as some of the
choices in the training procedure, the assumptions made on the zenith
angle and flavor dependence of the atmospheric fluxes, and the
uncertainty on the neutrino oscillation parameters.

\subsection{Impact of Training Choices}
\label{sec:traincho}

In order to verify the stability of the results with respect to the
minimization algorithm used in the neural network training procedure
first of all we have repeated the fit using genetic algorithms but
with a larger number of iterations ($N_{\rm it}$=300). The results are
shown in Fig.~\ref{fig:training}. We see that this increase in the
number of iterations does not lead to any substantial variation of the
allowed range of fluxes. This implies that the minimization had indeed
converged well before as it was illustrated in Fig.~\ref{fig:chi2gen}.

\FIGURE[t]{
  \includegraphics[width=0.65\textwidth]{training.epsi}
  \caption{\label{fig:training}%
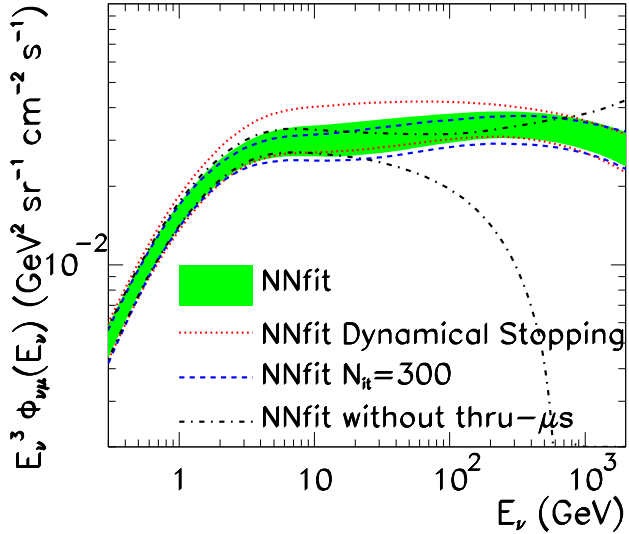
    Dependence of the allowed ranges of fluxes on different choices in
    the reference training. The full region is the result of our
    reference training. The dotted lines are the range of extracted
    fluxes if using dynamical stopping of the minimization with
    $\chi^2_{\mathrm{stop}}=160$. The dashed lines are the range of
    extracted fluxes if using genetic algorithms but with $N_{\rm
    it}=300$ iterations in the minimization. The dot-dashed lines are
    the range of extracted fluxes when removing the thrugoing muon
    sample from the fit.}%
  }

Second, we have repeated the fit by using dynamical stopping of the
training as alternative minimization strategy. In dynamical stopping,
instead of training each neural net a fixed number of iterations, the
training of each net is stopped independently when a certain condition
on the error function $E^{(k)}$ is satisfied. In particular, in this
training strategy one fixes a parameter $\chi^2_{\mathrm{stop}}$ and
then stops the fit to each replica separately when the condition
$E^{(k)} \le \chi^2_{\mathrm{stop}}$ is satisfied, with $E^{(k)}$ as
stated in Eq.~\eqref{eq:chi2}. It can be shown~\cite{f2ns} that the
typical values for $E^{(k)}$, are of the order of
$\chi^{2(0)}+N_{\dat}$. Therefore each different value of
$\chi^2_{\mathrm{stop}}$ will result in a different value of the total
$\chi^2_{\tot}$, Eq.~\eqref{eq:chi2tot}. Clearly one expects that the
higher the value of $\chi^2_{\tot}$ the larger the flux error ranges.
Thus to make the comparison meaningful we must chose a value of
$\chi^2_{\mathrm{stop}}$ which leads to a $\chi^2_{\tot}$ of
comparable value of the one obtained in the reference fit. We have
verified that for $\chi^2_{\mathrm{stop}}=160$, $\chi^2_{\tot}=76.5$
which is close enough to the reference fit value $\chi^2_{\tot}=74.6$.
The results for this alternative fit are shown in
Fig.~\ref{fig:training}. Again we see that the results obtained with
both minimization strategies are very compatible. The main effect of
using dynamical stopping is a slight increase of the allowed range of
fluxes in the intermediate energy region.

Finally, as a consistency check, we have verified that the obtained
energy dependence in a given energy range is mostly determined by the
data sample which is most sensitive to that energy range. We have done
so by repeating the reference fit but removing the thrugoing muon
sample in the analysis. The results are shown in
Fig.~\ref{fig:training} where we see that, while the fit is unaltered
at lower energies, the results get considerably different for
neutrinos with energies $E_\nu\gtrsim 50$ GeV which are the ones
responsible for thrugoing neutrino events as illustrated in
Fig.~\ref{fig:events}. Basically once the thrugoing muon sample is
removed from the fit, we have no experimental information for
$E_\nu\gtrsim 100$ GeV. Consequently the results of the fit at those
higher energies are just an unphysical extrapolation of the fit at
lower energies as clearly illustrated in the figure where we see that
there is no lower bound to the allowed flux at those higher energies.

This reflects the fact that the behavior of neural networks in the
extrapolation region is not determined by its behavior where more data
is available, as it would happen in fits with usual functional forms.
In principle, if the neural network had found some underlying physical
law which described the experimental data and which would be valid
both in the region where data is available and in the extrapolation
region, the value of the extrapolated fluxes could be less
unconstrained. However, this is not expected in this case since it is
precisely at energies of the order of $E_\nu \sim 100$ GeV that the
$\pi$ start interacting before they are able to decay which implies
that the underlying physical law is different below and above those
energies, and the extrapolation is therefore very much unconstrained.

\subsection{Impact of Choice for Zenith Angle and Flavor Flux Dependence}

As previously discussed, since the available experimental data from
Super-Kamiokande are not precise enough to allow for a simultaneous
independent determination of all the elements in the atmospheric
neutrino fluxes, we have restricted our analysis to the determination
of their energy dependence, and taken the zenith angle and flavor
dependence from a previous computation~\cite{honda}.

In order to assess the effects of this choice we first repeat the
reference fit but using as reference flux the Bartol
flux~\cite{bartol}. In the left panel of Fig.~\ref{fig:fluxcomp} we
compare the results of these two fits. We see that the results are
identical for $E_\nu \lesssim 10$ GeV as expected since both Honda and
Bartol calculations give very similar angular and flavour ratios at
those energies.  And for any energy the difference between the results
of both fits are much smaller than the differences between the
reference fluxes themselves, see Fig.~\ref{fig:fluxref}. This is what
is expected since the {\it flx} uncertainties included in both fits
represent the spread on the theoretical flux calculations of the
angular dependence and flavour ratios.

In the second column in Table~\ref{tab:resest} we give the statistical
estimators corresponding to the fit taking Bartol as a reference
flux. As expected the differences with results using Honda as
reference flux (first column in Table~\ref{tab:resest}) are very
small.

The importance of the choice of angular and flavour dependence can
also be addressed by studying the effect of removing from the analysis
the {\it flx} errors since those errors account for the spread of 
the predicted angular and flavour dependence among the different 
atmospheric flux calculations. In other words removing those errors
we force the neural net flux to follow the angular and flavour dependence
of the reference flux without allowing for any fluctuation about them.
The results of this fit and its comparison with
the reference fit are shown in the right panel of
Fig.~\ref{fig:fluxcomp}. The corresponding statistical estimators are
given in the third column of Table~\ref{tab:resest} while in
Fig.~\ref{fig:events_flx} we show the neural network predictions for
the event rates in this case.

\FIGURE[t]{
  \includegraphics[width=\textwidth]{fluxcomp.epsi}
  \caption{\label{fig:fluxcomp}%
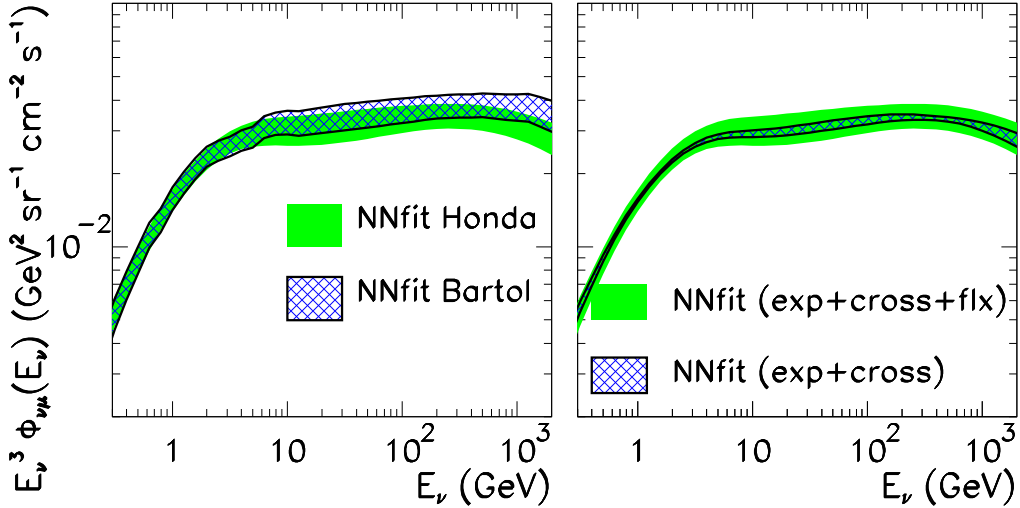
    {\bf left} Comparison of results with different reference fluxes.
    {\bf right} Comparison of determined fluxes in fits which include
    different sets of errors}%
  }

\FIGURE[t]{
  \includegraphics[width=\textwidth]{events_flx.epsi}
  \caption{\label{fig:events_flx}%
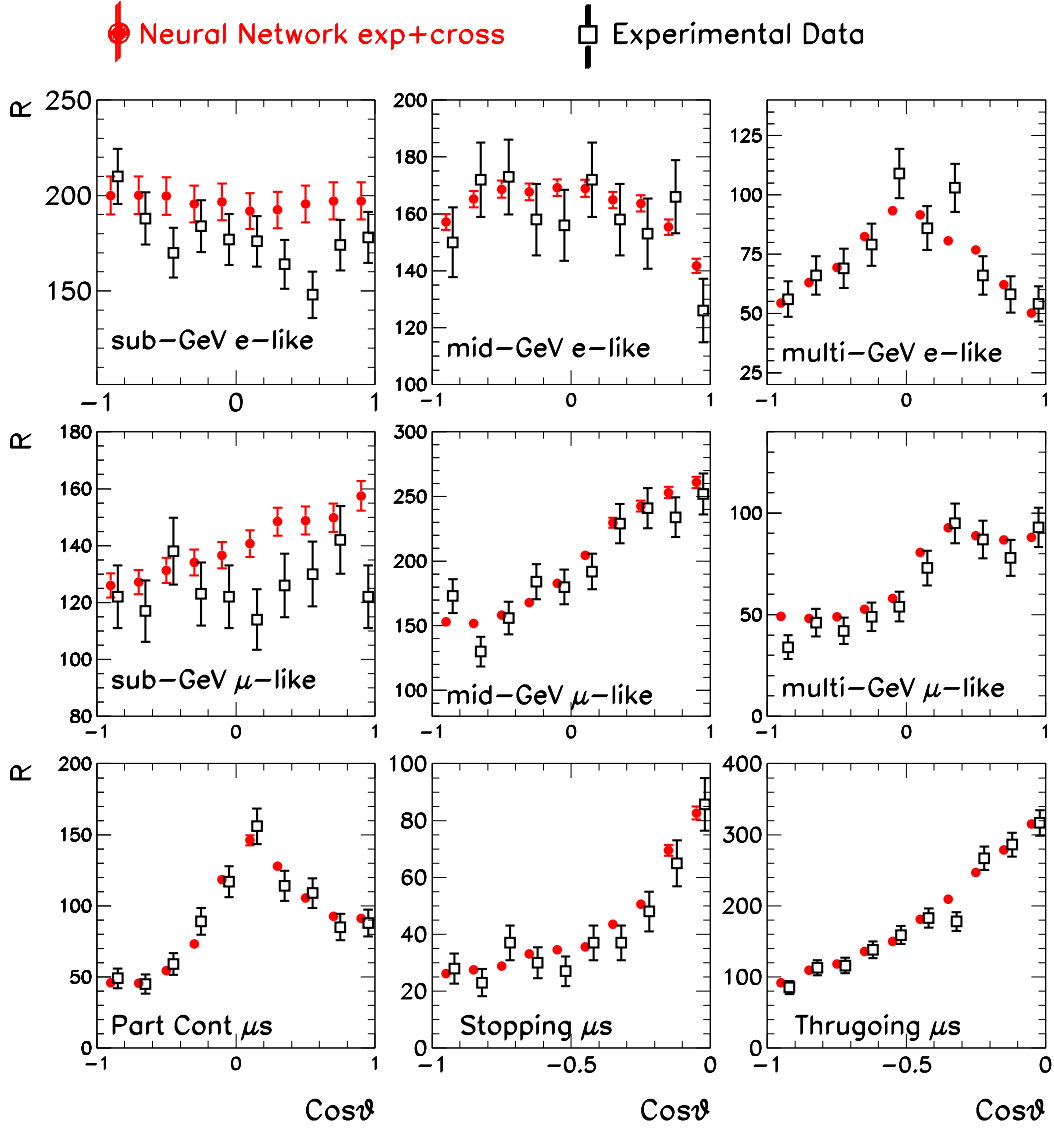
    Predicted number of atmospheric neutrino events using the
    atmospheric fluxes resulting from the fit with {\it exp+cross}
    errors only compared to the experimental data points.  The central
    values correspond to the average prediction and the error bars
    give the 1$\sigma$ ranges. Notice that only the statistical error
    is shown for the experimental data points.}%
  }

As expected, $\chi^2_{\rm tot}$ (see Table~\ref{tab:resest})
is larger once the {\it flx} errors
are not included in the fit, although given the small size of the {\it
flx} errors this increase is very moderate (only 1.1 units for 90 data
points). Equivalently from Table~\ref{tab:resest} we see the small impact
that the exclusion of the {\it flx} errors makes in the evaluation of the
statistical estimators of the experimental data 
$\langle \sigma^{\rm(exp)}\rangle_{\rm dat}$ ,
$\langle \rho^{\rm(exp)}\rangle_{\rm dat}$ , and
$\langle \cov^{\rm(exp)}\rangle_{\rm dat}$ which is again a reflection of
the small values of the {\it flx} compared to the experimental
statistical and systematic uncertainties.

There is, however, a much more important effect in the size of the
allowed range of fluxes and predicted rates in as seen in
Figs.~\ref{fig:fluxcomp} and~\ref{fig:events_flx}. As shown in
Fig.~\ref{fig:rel_err} the relative error of the flux is reduced by a
factor $\sim 3$--$4$ in the full energy range. At first this
considerable reduction of the relative flux error when removing only
the relatively small {\it flx} uncertainties may seem
counterintuitive. However this result is expected if the results of
the fit are consistent. This is because the zenith and flavor
dependence is not fitted from data.  As a consequence, if no
uncertainties are included associated with those, there are $N_E=20$
or $10$ binned rates of similar statistical weight (20 from the e-like
and mu-like distributions for sub- mid- and multi-GeV events and 10
for partially contained, stopping and thrugoing muons) contributing to
the determination of the flux in the same energy range with no allowed
fluctuations among them. Thus the associated uncertainty in the
determination of the energy dependence is indeed reduced by a factor
$\sqrt{N_E}\sim 3$--5. This is also reflected in
Table~\ref{tab:resest}, where we see that there is a reduction on the
statistical estimators which measure the average spread of the
predicted rates obtained when using the neural net fluxes, $\langle
\sigma^{\rm(net)}\rangle_{\rm dat}$ and $\langle
\cov^{\rm(net)}\rangle_{\rm dat}$.\footnote{Since $\langle
\cov^{\rm(net)}\rangle_{\rm dat}$ is proportional to the square of the
$\sigma^{\rm(net)}_i$, the reduction in this case is a factor $16\sim
(\sqrt{N_E})^2$.}

The inclusion of the {\it flx} uncertainties and their corresponding
pulls allows for the angular and flavour ratio of the fitted fluxes to
spread around their reference flux values. This results into an
effective {\it decoupling} of the contribution of the $N_E$ data
points to the fit at a given $E$ with the corresponding increase in
the flux relative error. Furthermore once the {\it flx} uncertainties
are included the range of the angular binned rate predictions are of
the same order of the statistical error of the experimental points
(see Fig.~\ref{fig:events}).

Finally let's also point out that, in general, adding or removing some
source of uncertainty does not only change the size of the associated
errors in the parametrization but also the position of the minimum,
that is, the features of the {\it best-fit} flux, although this effect
is small in the present case.

\subsection{Impact of Oscillation Parameters}

So far the results presented for the different fits have been done
assuming $\nu_\mu\rightarrow \nu_\tau$ oscillations with oscillation
parameters fixed to
\begin{equation}
    \sin^2 2\theta_{\mathrm{atm}}=1, \qquad  \Delta m_{\mathrm{atm}}^2=
    2.2~\times~10^{-3}~ \mathrm{eV}^2 \,.
\end{equation}
In order to address the impact on the results of the assumed value of
the oscillation parameters we repeat the fit with these parameters
varying within their 1-sigma ranges as allowed by global fits to
neutrino oscillations data. In particular we consider the range
\begin{equation} \label{eq:oscrange}
    \sin^2 2\theta_{\mathrm{atm}}\ge 0.96,
    \,, \qquad  1.8~\times~10^{-3}~ \mathrm{eV}^2  \le
    \Delta m_{\mathrm{atm}}^2\le
    2.7~\times~10^{-3} ~\mathrm{eV}^2 \,.
\end{equation}
The results are shown in Fig.~\ref{fig:oscplot} where we show the
envelope of the results obtained varying each time one of the neutrino
oscillation parameters. As we can see in Fig.~\ref{fig:oscplot}, the
contribution to the total error from the uncertainty in the neutrino
oscillation parameters is rather small. Therefore we can be confident
than the impact in our results of the uncertainties in the oscillation
parameters is very small, and moreover this uncertainty can be
systematically reduced as our knowledge of neutrino oscillation
parameters increases.

\FIGURE[t]{
  \includegraphics[width=0.70\textwidth]{oscplot.epsi}
  \caption{\label{fig:oscplot}%
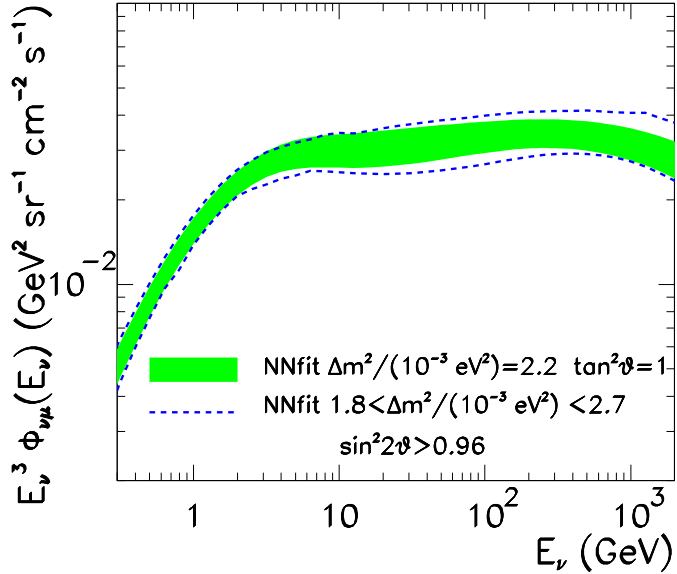
    Comparison of the reference fit with the envelope of fits obtained
    varying the atmospheric neutrino oscillation parameters given in
    the label.}%
  }

\section{Summary and Outlook}
\label{sec:conclu}

In this work we have presented the first results on the determination
of the energy dependence of the atmospheric neutrino fluxes from the
data on atmospheric neutrino event rates measured by the
Super-Kamiokande experiment. In order to bypass the problem of the
unknown functional form for the neutrino fluxes we have made use of
artificial neural networks as unbiased interpolants. On top of this, a
faithful estimation of the uncertainties of the neutrino flux
parametrization has been obtained by the use of Monte Carlo methods.

In our analysis we have relied on the zenith and flavour dependence of
the flux as predicted by some of the atmospheric flux calculations in
Refs.~\cite{honda,bartol,others}. Also, the fluxes are determined
under the assumption that oscillation parameters will eventually be
independently determined by non atmospheric neutrino experiments with
a value close to the present best fit. We have estimated the
uncertainties associate with these choices by performing alternative
fits to the data where some of these assumptions were changed and/or
relaxed. 

Our main result is presented in Fig.~\ref{fig:fluxref}.  We have found
that until about $E_\nu\sim 1$ TeV we have a good understanding of the
normalization of the fluxes and the present accuracy from
Super-Kamiokande neutrino data is comparable with the theoretical
uncertainties from the numerical calculations.  The results of our
alternative fits shows that if one assumes that the present
uncertainties of the angular dependence have been properly estimated,
it turns out that the assumed angular dependence has very little
effect on the determination of the energy dependence of the fluxes.
Thus the determined atmospheric neutrino fluxes could be used as an
alternative of the existing flux calculations, and are available upon
request to the authors.

\FIGURE[t]{
  \includegraphics[width=0.70\textwidth]{refit_extra.epsi}
  \caption{\label{fig:fluxref_extra}%
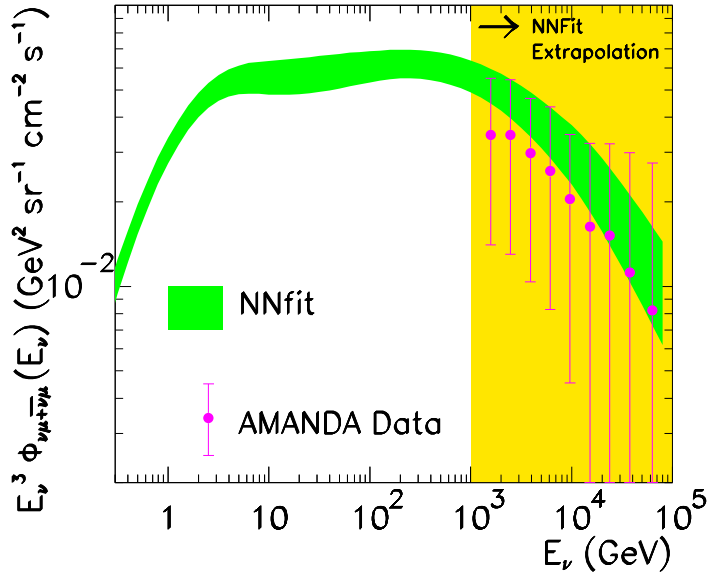
    Results for the reference fit for the angular averaged muon
    neutrino plus anti neutrino flux extrapolated to the high energy
    region compared to the corresponding data from
    AMANDA~\cite{icecubedata}.}%
  }

The results of this work can be extended in several directions.  It
would be interesting to include in the analysis the atmospheric
neutrino data from detectors that probe the high energy region, like
AMANDA~\cite{amandadata,icecubedata} or ICECUBE~\cite{icecube}. To
illustrate the reach of the presently available statistics at those
energies we show in Fig.~\ref{fig:fluxref_extra} the results for the
reference fit for the angular averaged muon neutrino plus antineutrino
flux extrapolated to the high energy region compared to the data from
AMANDA~\cite{amandadata,icecubedata}. Notice that, as mentioned above,
the behavior of neural networks in the extrapolation region is not
determined by its behavior where data is available, as it would happen
in fits with usual functional forms.  As a consequence the values of
the extracted fluxes in the extrapolation region can be extremely
unphysical as described in Sec.~\ref{sec:traincho}. To improve the
extrapolation, one could use high-energy functional forms for the
atmospheric neutrino flux, for example those presented
in~\cite{gaisser}, which have been used in~\cite{mcfit} to fit
analytical expressions for the fluxes to the Monte Carlo simulations.
The implementation of this strategy is postponed to future work, when
also the AMANDA and ICECUBE atmospheric neutrino data will be
incorporated in the fit.

Furthermore one could assess the effects of determining from
experimental data the full energy, zenith and flavor dependence of the
atmospheric neutrino fluxes together with the oscillation parameters,
in particular in the context of the expected data from future megaton
neutrino detectors~\cite{hyperk,uno}.

\acknowledgments

We thank F.~Halzen for comments and careful reading of the manuscript.
J.~R.~would like to acknowledge the members of the NNPDF
Collaboration: Luigi del Debbio, Stefano Forte, Jose Ignacio Latorre
and Andrea Piccione, since a sizable part of this work is related to
an upcoming common publication~\cite{pdf2}. M.C.~G-G would like to
thank the CERN theory division for their hospitality during the weeks
previous to the finalization of this work. This work is supported by
National Science Foundation grant PHY-0354776 and by Spanish Grants
FPA-2004-00996 and AP2002-2415.

\appendix

\section{Statistical Estimators}
\label{sec:estimators}

In this Appendix we summarize the estimators used to validate the
generation of the Monte Carlo sample of replicas of the experimental
data. The corresponding estimators that validate the neural network
training can be straightforwardly obtained by replacing the
superscript $~^{(\art)}$ with $~^{(\net)}$.

\begin{itemize}
  \item Average over the number of replicas for each experimental point $i$
    \begin{equation}
        \la
        R_i^{(\art)}\ra_{\rep}=\frac{1}{N_{\rep}}\sum_{k=1}^{N_{\rep}}
        R_i^{(\art)(k)}\,.
    \end{equation}
    
  \item Associated variance
    \begin{equation} \label{eq:var}
        \sigma_i^{(\art)}=\sqrt{\la\lp R_i^{(\art)}\rp^2\ra_{\rep}-
          \la R_i^{(\art)}\ra^2_{\rep}} \,.
    \end{equation}

  \item Associated covariance
    \begin{gather}
        \label{eq:rho}
        \rho_{ij}^{(\art)}=\frac{\la R_i^{(\art)}R_j^{(\art)}\ra_{\rep}-
          \la R_i^{(\art)}\ra_{\rep}\la R_j^{(\art)}\ra_{\rep}}{\sigma_i^{(\art)}
          \sigma_j^{(\art)}} \,,
        \\
        \label{eq:cov}
        \mathrm{cov}_{ij}^{(\art)}=\rho_{ij}^{(\art)}\sigma_i^{(\art)}
        \sigma_j^{(\art)} \,.
    \end{gather}
    The three above quantities provide the estimators of the
    experimental central values, errors and correlations
    which one extracts from the sample
    of experimental data.
    
  \item Mean variance and percentage error on central values
    over the $N_{\dat}$ data points.
    \begin{align}
        \la V\lc\la R^{(\art)}\ra_{\rep}\rc\ra_{\dat} &=
        \frac{1}{N_{\dat}}\sum_{i=1}^{N_{\dat}}\lp \la R_i^{(\art)}\ra_{\rep}-
        R_i^{(\mrexp)}\rp^2 \,,
        \\
        \la PE\lc\la R^{(\art)}\ra_{\rep}\rc\ra_{\dat} &=
        \frac{1}{N_{\dat}}\sum_{i=1}^{N_{\dat}}\lc\frac{ \la R_i^{(\art)}\ra_{\rep}-
          R_i^{(\mrexp)}}{R_i^{(\mrexp)}}\rc \,.
    \end{align}
    We define analogously $\la V\lc\la
    \sigma^{(\art)}\ra_{\rep}\rc\ra_{\dat}$, $\la V\lc\la
    \rho^{(\art)}\ra_{\rep}\rc\ra_{\dat}$, $\la V\lc\la
    \mathrm{cov}^{(\art)}\ra_{\rep}\rc\ra_{\dat}$ and $\la PE\lc\la
    \sigma^{(\art)}\ra_{\rep}\rc\ra_{\dat}$, $\la PE\lc\la
    \rho^{(\art)}\ra_{\rep}\rc\ra_{\dat}$ and $\la PE\lc\la
    \mathrm{cov}^{(\art)}\ra_{\rep}\rc\ra_{\dat}$, for errors,
    correlations and covariances respectively.
    
    These estimators indicate how close the averages over generated
    data are to the experimental values. Note that in averages over
    correlations and covariances one has to use the fact that
    correlation and covariances matrices are positive definite, and
    thus one has to be careful to avoid double counting. For example,
    the percentage error on the correlation will be defined as
    \begin{equation}
        \la PE\lc\la
        \rho^{(\art)}\ra_{\rep}\rc\ra_{\dat}=\frac{2}{N_{\dat}\lp
          N_{\dat}+1\rp}\sum_{i=1}^{N_{\dat}}
        \sum_{j=i}^{N_{\dat}}\lc\frac{ \la \rho_{ij}^{(\art)}\ra_{\rep}-
          \rho_{ij}^{(\mrexp)}}{\rho_{ij}^{(\mrexp)}}\rc \,,
    \end{equation}
    and similarly for averages over elements of the covariance matrix.
    
  \item Scatter correlation:
    \begin{equation}
        r\lc R^{(\art)}\rc=\frac{\la R^{(\mrexp)}\la R^{(\art)}
          \ra_{\rep}\ra_{\dat}-\la R^{(\mrexp)}\ra_{\dat}\la\la R^{(\art)}
          \ra_{\rep}\ra_{\dat}}{\sigma_s^{(\mrexp)}\sigma_s^{(\art)}} \,
    \end{equation}
    where the scatter variances are defined as
    \begin{align}
        \sigma_s^{(\mrexp)} &=
        \sqrt{\la \lp R^{(\exp)}\rp^2\ra_{\dat}-
          \lp \la  R^{(\exp)}\ra_{\dat}\rp^2} \,,
        \\
        \sigma_s^{(\art)} &=
        \sqrt{\la \lp \la R^{(\art)}\ra_{\rep}\rp^2\ra_{\dat}-
          \lp \la  \la R^{(\art)}\ra_{\rep} \ra_{\dat}\rp^2} \,.
    \end{align}
    We define analogously $r\lc\sigma^{(\art)}\rc$,
    $r\lc\rho^{(\art)}\rc$ and $r\lc\mathrm{cov}^{(\art)}\rc$. Note
    that the scatter correlation and scatter variance are not related
    to the variance and correlation
    Eqs.~\eqref{eq:var}-\eqref{eq:cov}. The scatter correlation
    indicates the size of the spread of data around a straight line.
    Specifically $r\lc \sigma^{(\net)}\rc=1$ implies that $\la
    \sigma_i^{(\net)}\ra$ is proportional to $\sigma_i^{(\mrexp)}$.
    
  \item Average variance:
    \begin{equation} \label{eq:avvar}
        \la \sigma^{(\art)}\ra_{\dat}=\frac{1}{N_{\dat}}
        \sum_{i=1}^{N_{\dat}}\sigma_i^{(\art)} \,.
    \end{equation}
    We define analogously $\la\rho^{(\art)}\ra_{\dat}$ and
    $\la\mathrm{cov}^{(\art)}\ra_{\dat}$, as well as the corresponding
    experimental quantities. These quantities are interesting because
    even if $r$ are close to 1 there could still be a systematic bias
    in the estimators Eqs.~\eqref{eq:var}-\eqref{eq:cov}. This is so
    since even if all scatter correlations are very close to 1, it
    could be that some of the Eqs.~\eqref{eq:var}-\eqref{eq:cov} where
    sizably smaller than its experimental counterparts, even if being
    proportional to them.
\end{itemize}

The typical scaling of the various quantities with the number of
generated replicas $N_{\rep}$ follows the standard behavior of
gaussian Monte Carlo samples. For instance, variances on central
values scale as $1/N_{\rep}$, while variances on errors scale as
$1/\sqrt{N_{\rep}}$. Also, because
\begin{equation}
    V[\rho^{(\art)}] 
    = \frac{1}{N_{\rep}}\lc1-\lp\rho^{(\exp)}\rp^2\rc^2,
\end{equation}
the estimated correlation fluctuates more for small values of
$\rho^{(\exp)}$, and thus the average correlation tends to be larger
than the corresponding experimental value.

\section{The Monte Carlo Approach to Error Estimation}
\label{sec:mcexample}

In this Appendix we show with a simple example how the Monte Carlo
approach to error estimation is equivalent to the standard approach,
based on the condition $\Delta\chi^2=1$ for the determination of
confidence levels, with the assumption of gaussian errors, up to
linearized approximations. For a more detailed analysis of this
statistical technique the reader is referred to~\cite{cowan}.

Let us consider two pairs of independent measurements of the same
quantity, $x_1\pm \sigma_1$ and $x_2\pm \sigma_2$ with gaussian
uncertainties. The distribution of true values of the variable $x$ is
a gaussian distribution centered at
\begin{equation}
    \overline{x}=\frac{x_1\sigma_2^2+x_1\sigma_2^2}{\sigma_1^2+\sigma_2^2} \,,
\end{equation}
and with variance determined by the $\Delta\chi^2=1$ tolerance
criterion,
\begin{equation} \label{eq:varmc}
    \sigma^2=\frac{\sigma_1^2\sigma_2^2}{\sigma_1^2+\sigma_2^2}\,.
\end{equation}
To obtain the above results, note that if errors are gaussianly
distributed, the maximum likelihood condition implies that the mean
$\overline{x}$ minimizes the $\chi^2$ function
\begin{equation}
    \chi^2=\frac{\lp x_1-\overline{x}\rp}{\sigma_1^2}+
    \frac{\lp x_2-\overline{x}\rp}{\sigma_2^2}  \,,
\end{equation}
and the variance $\sigma$ is determined by the condition
\begin{equation}
    \Delta\chi^2=\chi^2\lp \overline{x}+\sigma\rp-
    \chi^2\lp \overline{x}\rp \,,
\end{equation}
which for $\Delta\chi^2=1$ leads to Eq.~\eqref{eq:varmc}. Note that
these properties only hold for gaussian measurements.

An alternative way to compute the mean and the variance of the
combined measurements $x_1$ and $x_2$ is the Monte Carlo method:
generate $N_{\rep}$ replicas of the pair of values $x_1,x_2$
gaussianly distributed with the appropriate error,
\begin{gather}
    \label{eq:rep11}
    x_1^{(k)}=x_1+r_1^{(k)}\sigma_1, \qquad k=1,\ldots,N_{\rep} \,,
    \\[1mm]
    \label{eq:rep22}
    x_2^{(k)}=x_2+r_2^{(k)}\sigma_2, \qquad k=1,\ldots,N_{\rep} \,,
\end{gather}
where $r^{(k)}$ are univariate gaussian random numbers. One can then
show that for each pair, the weighted average
\begin{equation}
    \overline{x}^{(k)}=\frac{x_1^{(k)}\sigma_2^2+
      x_1^{(k)}\sigma_2^2}{\sigma_1^2+\sigma_2^2} \,,
\end{equation}
is gaussianly distributed with central value and width equal to the
one determined in the previous case. That is, it can be show that for
a large enough value of $N_{\rep}$,
\begin{equation}
    \la \overline{x}^{(k)}\ra_{\rep}=\frac{1}{N_{\rep}}
    \sum_{k=1}^{N_{\rep}}\overline{x}^{(k)}=\overline{x} \,,
\end{equation}
and for the variance
\begin{equation}
    \sigma^2=\la  \lp\overline{x}^{(k)}\rp^2\ra_{\rep}-
    \la  \overline{x}^{(k)}\ra_{\rep}^2=\frac{\sigma_1^2\sigma_2^2}{
      \sigma_1^2+\sigma_2^2} \,,
\end{equation}
which is the same result, Eq.~\eqref{eq:varmc}, as obtained from the
$\Delta\chi^2=1$ criterion. This shows that the two procedures are
equivalent in this simple case.

The generalization to $N_{\dat}$ gaussian correlated measurements is
straightforward. Let us consider for instance that the two
measurements $x_1$ and $x_2$ are not independent, but that they have
correlation $\rho_{12} \le 1$. To take correlations into account, one
uses the same Eqs.~\eqref{eq:rep11} and \eqref{eq:rep22} to generate
the sample of replicas of the measurements, but this time the random
numbers $r_1^{(k)}$ and $r_2^{(k)}$ are univariate gaussian correlated
random numbers, that is, they satisfy
\begin{equation}
    \la r_1r_2\ra_{\rep}=\frac{1}{N_{\rep}}\sum_{k=1}^{N_{\rep}}
    r_1^{(k)}r_2^{(k)}=\rho_{12} \,.
\end{equation}
With this modification, the sample of Monte Carlo replicas of $x_1$
and $x_2$ also reproduces the experimental correlations. This can be
seen with the standard definition of the correlation,
\begin{equation}
    \rho\equiv \la \frac{ \lp x_1^{(k)}-x_1\rp
      \lp x_2^{(k)}-x_2\rp}{\sigma_1\sigma_2}\ra_{\rep}=\la r_1r_2\ra_{\rep}=
    \rho_{12} \,.
\end{equation}
Therefore, the Monte Carlo approach also correctly takes into account
the effects of correlations between measurements.

In realistic cases, the two procedures are equivalent only up to
linearizations of the underlying law which describes the experimental
data. We take the Monte Carlo procedure to be more faithful in that it
does not involve linearizing the underlying law in terms of the
parameters. Note that as emphasized before, the error estimation
technique that is described in this thesis does not depend on whether
one uses neural networks or polynomials as interpolants.

\section{Neural Network Details}
\label{sec:neuralnetdet}

Artificial neural networks~\cite{nnbook,nnthesis} provide unbiased
robust universal approximates to incomplete or noisy data. An
artificial neural network consists of a set of interconnected units
({\it neurons}). The {\it activation} state $\xi_i^{(l)}$ of a neuron
is determined as a function of the activation states of the neurons
connected to it. Each pair of neurons $(i,j)$ is connected by a
synapses, characterized by a {\it weight} $\omega_{ij}$. In this work
we will consider only multilayer feed-forward neural networks. These
neural networks are organized in ordered layers whose neurons only
receive input from a previous layer. In this case the activation state
of a neuron in the (l+1)-th layer is given by
\begin{gather}
    \xi^{(l+1)}_i = g\lp h^{(l+1)}_i\rp \,,
    \qquad i=1,\ldots,n_{l+1} \,, \qquad l=1,\ldots, L-1 \,,
    \\
    \label{eq:hfun}
    h^{(l+1)}_i= \sum_{j=1}^{n(l)}\omega_{ij}^{(l)}\xi^{(l)}_j+
    \theta_{i}^{(l+1)} \,,
\end{gather}
where $\theta_{i}^{(l)}$ is the {\it activation threshold} of the
given neuron, $n_l$ is the number of neurons in the l-th layer, $L$ is
the number of layers that the neural network has, and $g(x)$ is the
{\it activation function} of the neuron, which we will take to be a
sigma function,
\begin{equation}
    g(x)=\frac{1}{1+e^{-x}} \,,
\end{equation}
except in the last layer, where we use a linear activation function
$g(x)$. This enhances the sensitivity of the neural network, avoiding
the saturation of the neurons in the last layer. The fact that the
activation function $g(x)$ is non-linear allows the neural network to
reproduce nontrivial functions.

Therefore multilayer feed-forward neural networks can be viewed as
functions $F:\mathcal{R}^{n_1}\to \mathcal{R}^{n_L}$ parametrized by
weights, thresholds and activation functions,
\begin{equation}
    \xi^{L}_j = F\lc \xi^{(1)}_i,\omega_{ij}^{(l)},\theta_i^{(l)},g\rc ,
    \qquad
    j = 1\ldots,n_L\,.
\end{equation}
It can be proved that any continuous function, no matter how complex,
can be represented by a multilayer feed-forward neural network. In
particular, it can be shown~\cite{nnbook,nnthesis} that two hidden
layers suffice for representing an arbitrary function.

In our particular case the architecture of the neural network used is
1-5-5-1, which means that it has a single input neuron (whose value is
the neutrino energy), two hidden layers with 5 neurons each and a
final output neuron (whose value is the atmospheric neutrino flux).

\section{Genetic Algorithms}
\label{sec:ga}

Genetic algorithms are the generic name of function optimization
algorithms that do not suffer of the drawbacks that deterministic
minimization strategies have when applied to problems with a large
parameter space.  This method is specially suitable for finding the
global minima of highly nonlinear problems. 

All the power of genetic algorithms lies in the repeated application
of two basic operations: mutation and selection.
The first step is to encode the information of the parameter space of
the function we want to minimize into an ordered chain, called {\it
chromosome}. If $N_{\parr}$ is the size of the parameter space, then a
point in this parameter space will be represented by a chromosome,
\begin{equation} \label{chain}
    {\bf a}=\lp a_1,a_2,a_3,\ldots,a_{N_{\parr}}\rp \,.
\end{equation}
In our case each {\it bit} $a_i$ of a chromosome corresponds to either
a weight $\omega^{(l)}_{ij}$ or a threshold $\theta^{(l)}_i$ of a
neural network. Once we have the parameters of the neural network
written as a chromosome, we replicate that chain until we have a
number $N_{\tot}$ of chromosomes.  Each chromosome has an associated
{\it fitness} $f$, which is a measure of how close it is to the best
possible chromosome (the solution of the minimization problem under
consideration). In our case, the fitness of a chromosome is given by
the inverse of the function to minimize, $ E^{k}$ given in 
Eq.~\eqref{eq:chi2}.

Then we apply the two basic operations:
\begin{itemize}
  \item Mutation:
    Select randomly a {\it bit} (an element of the chromosome) and
    {\it mutate} it. The size of the mutation is called {\it mutation
    rate} $\eta$, and if the k-th bit has been selected, the mutation
    is implemented as
    \begin{equation}
	a_k \to a_k + \eta \lp r-\frac{1}{2} \rp \,,
    \end{equation}
    where $r$ is a uniform random number between 0 and 1. 
    The optimal size of the mutation rate must be determined for each
    particular problem, or it can be adjusted dynamically as a
    function of the number of iterations.
  \item Selection:
    Once mutations and crossover have been performed into the
    population of individuals characterized by chromosomes, the
    selection operation ensures that individuals with best fitness
    propagate into the next generation of genetic algorithms. Several
    selection operators can be used. The simplest method is to select
    simply the $N_{\mathrm{chain}}$ chromosomes, out of the total
    population of $N_{\mathrm{tot}}$ individuals, with best fitness. 
\end{itemize}

The procedure is repeated iteratively until a suitable convergence
criterion is satisfied. Each iteration of the procedure is called a
{\it generation}. A general feature of genetic algorithms is that the 
fitness approaches
the optimal value within a relatively small number of generations, as
seen in Fig.~\ref{fig:chi2gen} for the present problem.

\end{document}